\documentclass[pdflatex,sn-mathphys-num,dvipsnames]{sn-jnl}%om
\usepackage{graphicx}%
\usepackage{multirow}%
\usepackage{amsmath,amssymb,amsfonts}%
\usepackage{amsthm}%
\usepackage{mathrsfs}%
\usepackage[title]{appendix}%
\usepackage{textcomp}%
\usepackage{manyfoot}%
\usepackage{booktabs}%
\usepackage{algorithm}%
\usepackage{algorithmicx}%
\usepackage{algpseudocode}%
\usepackage{listings}%
\theoremstyle{thmstyleone}%
\usepackage{adjustbox}
\usepackage{array}
\usepackage{pdflscape}
\usepackage{geometry}
\usepackage{wrapfig,lipsum,booktabs}
\usepackage{lineno}
\usepackage{mdframed}
\usepackage{pdflscape}
\theoremstyle{thmstyletwo}%
\theoremstyle{thmstylethree}%
\usepackage{hyperref}

\usepackage[nomargin,inline,draft,author=]{fixme} 
\fxusetheme{color}
\definecolor{fxwarning}{rgb}{0.8,0.0000,0.0000}
\FXRegisterAuthor{hm}{hdm}{hm}

\usepackage{soul} % for the command \hl
\begin{document}
\title{Programmable 200 GOPS Hopfield-inspired photonic Ising machine}

\author[1,3]{\fnm{Nayem} \sur{AL-Kayed}}\email{nayemal.kayed@queensu.ca}
\equalcont{These authors contributed equally to this work.}

\author[2,3]{\fnm{Charles \sur{St-Arnault}}}\email{charles.st-arnault@mail.mcgill.ca}
\equalcont{These authors contributed equally to this work.}

\author[1,3]{\fnm{Hugh} \sur{Morison}}\email{hugh.morison@queensu.ca}
\equalcont{These authors contributed equally to this work.}

\author[1,3,4,6]{\fnm{A.} \sur{Aadhi}}\email{aadhi.a@queensu.ca}

\author[5]{\fnm{Chaoran} \sur{Huang}}\email{crhuang@ee.cuhk.edu.hk}

\author[4]{\fnm{Alexander N.} \sur{Tait}}\email{alex.tait@queensu.ca}

\author[2]{\fnm{David V.} \sur{Plant}}\email{david.plant@mcgill.ca}

\author[1,3,4]{\fnm{Bhavin J.} \sur{Shastri}}\email{shastri@ieee.org}

\affil[1]{\orgdiv{Centre for Nanophotonics, Department of Physics, Engineering Physics, and Astronomy}, \orgname{Queen's University}, 
\city{Kingston}, \postcode{K7L~3N6}, \state{ON}, \country{Canada}}

\affil[2]{\orgdiv{Department of Electrical and Computer Engineering}, \orgname{McGill University},  
\city{Montreal}, \postcode{H3A~2A7}, \state{QC}, \country{Canada}}

\affil[3]{\orgname{Milkshake Technology Inc.}, \city{Kingston}, \postcode{K7M 6R4}, \state{ON}, \country{Canada}}

\affil[4]{\orgdiv{Smith Engineering, Department of Electrical and Computer Engineering}, \orgname{Queen's University}, \city{Kingston}, \postcode{K7L~3N6}, \state{ON}, \country{Canada}}

\affil[5]{\orgdiv{Department of Electronic Engineering}, \orgname{The Chinese University of Hong Kong}, \city{Shatin}, \state{New Territories}, \country{Hong Kong}}

\affil[6]{\orgdiv{Optics and Photonics Centre}, \orgname{Indian Institute of Technology Delhi}, \city{Delhi 110016}, \country{India}}

\abstract{
\textbf{Ising machines offer a compelling approach to addressing NP-hard problems~\cite{mohseni2022ising}, but physical realizations that are simultaneously scalable, reconfigurable, fast, and stable remain elusive. Quantum annealers, like D-Wave's cryogenic hardware, target combinatorial optimization tasks, but quadratic scaling of qubit requirements with problem size limits their scalability on dense graphs~\cite{hamerly2019experimental}. Here, we introduce a programmable, stable, room-temperature optoelectronic oscillator (OEO)-based Ising machine with linear scaling in spin representation. Inspired by Hopfield networks~\cite{hopfield1982}, our architecture solves fully-connected problems with up to 256 spins (65,536 couplings), and $>$41,000 spins (205,000+ couplings) if sparse. Our system leverages cascaded thin-film lithium niobate modulators, a semiconductor optical amplifier, and a digital signal processing (DSP) engine in a recurrent time-encoded loop, demonstrating potential $>$200 giga-operations per second for spin coupling and nonlinearity. This platform achieves the largest spin configuration in an OEO-based photonic Ising machine, enabled by high intrinsic speed. We experimentally demonstrate best-in-class solution quality for Max-Cut problems of arbitrary graph topologies (2,000 and 20,000 spins) among photonic Ising machines and obtain ground-state solutions for number partitioning~\cite{lucas2014ising} and lattice protein folding~\cite{dill2012protein}---benchmarks previously unaddressed by photonic systems. Our system leverages inherent noise from high baud rates to escape local minima and accelerate convergence. Finally, we show that embedding DSP---traditionally used in optical communications---within optical computation enhances convergence and solution quality, opening new frontiers in scalable, ultrafast computing for optimization, neuromorphic processing, and analog AI.} }

%\keywords{Ising Model, Hopfield Neural Network, Optical Computing, Integrated Photonics, Thin-film Lithium Niobate, Combinatorial Optimization, Number Partitioning, Protein Folding.}
\maketitle

\section{Introduction}

\label{sec:intro}
The century-old Ising model~\cite{ising1924beitrag}, originally developed to study the collective behavior of magnetic spins, has found broad applications across various fields, from solid-state and statistical physics to quantum information, structural biology, and beyond~\cite{budrikis}. Complex problems such as protein folding, number partitioning, and prime factorization belong to the class of nondeterministic polynomial (NP) problems, which remain computationally intractable even for modern supercomputers~\cite{garey1979}. Fortunately, many of these problems can be mapped onto an Ising Hamiltonian, allowing specialized hardware to emulate Ising dynamics and efficiently approximate, or even exactly identify optimal solutions. A wide variety of Ising machines have been proposed, including those based on trapped ions, digital electronics, memristors, superconducting circuits, spintronics, and photonics~\cite{mohseni2022ising}. Among these, photonic Ising machines have garnered significant interest due to their potential for high-speed operation, massive parallelism via multiplexing, and energy efficiency~\cite{mcmahon2023physics,shastri2021photonics,ZhangReview2025,gao2024photonic}. However, simultaneously achieving scalability, speed, and reconfigurability remains an unresolved challenge. Here, we demonstrate a Hopfield-inspired programmable photonic Ising machine capable of supporting up to 256 all-to-all connected spins (65,536 spin couplings) or 41,209 sparsely connected spins (205,233 couplings). For the largest problem considered, the system achieves a feedforward latency of $\sim$4.1~$\mu$s and performs over 200~giga-operations per second (GOPS). This represents the largest spin configuration demonstrated to date in an optoelectronic oscillator (OEO)-based Ising machine, enabled by the system's high intrinsic photonic hardware speed (see Table~1). 

\begin{figure}[p]
\centering
\includegraphics[width=.7\paperwidth,height=.9\paperheight,keepaspectratio]{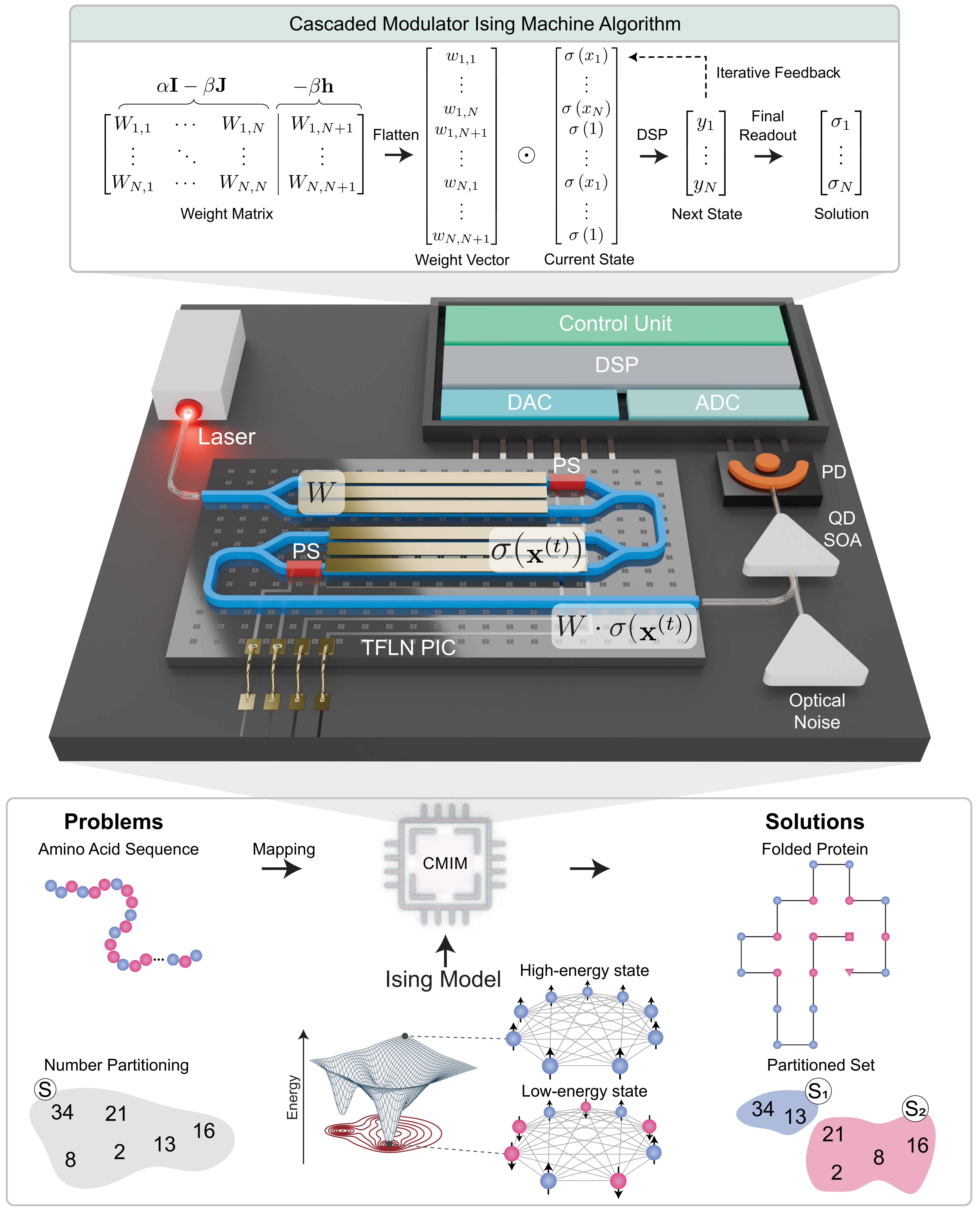}
\caption{\textbf{Conceptual illustration of the proposed CMIM.} Mapping combinatorial optimization problems, such as number partitioning and protein folding, to the Ising model provides an efficient framework for addressing NP-complexity challenges. The minimum energy configurations in the energy landscape ($H$) correspond to solutions to the problem. The hardware comprises two cascaded thin-film lithium niobate (TFLN) modulators~\cite{lin2024}, which encode the input spin vectors and the weight matrix (\textbf{W}). A quantum-dot semiconductor optical amplifier (QD SOA) amplifies the element-wise multiplied signals. A bulk SOA serves as a controlled optical noise source to accelerate convergence when the system struggles to find optimum solutions due to frustrated energy landscapes containing multiple local minima. The optical signal is then converted to an electrical signal using a photodetector (PD). The digital-to-analog converter (DAC) encodes the input signals, while the signal from the PD is sampled by an analog-to-digital converter (ADC), forming an electro-optic (EO) feedback system. Digital signal processing (DSP) stacks process the input and output signals at each iteration. The feedback loop is repeated across multiple iterations until convergence to the ground state solution is achieved. The mathematical representation (top) shows how the weight matrix is determined from the fixed coupling matrix $\mathbf{J}$ and the local bias $\mathbf{h}$ corresponding to the given problem. The coupling matrix is flattened for element-wise multiplication, and the resulting values are summed to perform matrix-vector multiplication. The DSP processes the received data and routes it back through the feedback path. Over successive iterations, the system converges to the ground state solution ($\sigma$).}
\label{fig:cmim}
\end{figure}

Hopfield neural networks have long attracted interest across fields such as control systems, game theory, and bioinformatics~\cite{hopfield1982}. These networks exhibit key features including associative memory, energy-based dynamics, and simple, flexible connectivity. Our photonic cascaded modulator Ising machine (CMIM; depicted in Fig.~\ref{fig:cmim}) emulates Hopfield-like dynamics \cite{kalinin2022} through a simple time-encoded recurrent architecture incorporating cascaded thin-film lithium niobate (TFLN) modulators~\cite{lin2024}, a quantum-dot semiconductor optical amplifier (QD SOA), and a digital signal processing (DSP) engine, all driven by a 256~GSa/s test and measurement setup. The large bandwidth (110~GHz), low DC V$_{\pi}$ (1.5~V), and  (sinusoidal) electro-optic nonlinearity of the TFLN modulators enable a stable, high baud rate realization of large-scale photonic Ising spin networks. Operating at a high baud rate allows for shorter time-encoded spin intervals, increasing the number of spins while reducing time-to-solution (TTS), which is crucial for scalable systems. However, this also introduces challenges such as signal degradation, synchronization errors, and limited bit resolution, all of which can degrade solution quality (i.e., proximity to the ground state). A previous attempt showed a low-speed Ising machine using optical modulators operating at 6.25~GBaud and solving only basic benchmark problems~\cite{li2024scalable}. Our Ising machine leverages DSP and a controllable optical noise source to escape local minima and solve large-scale problems at 106~GBaud. While DSP is traditionally used in telecommunications for high-speed, long-distance transmission, we discovered that embedding it in a photonic computing engine is not only essential for stable operation at high computational speeds but also accelerates convergence and improves solution quality. Moreover, our physical system naturally benefits from inherent noise from high baud rates, which enhances the annealing process by helping the system escape local minima. We believe this convergence of computing and communication principles can open new frontiers in photonic information processing systems, including programmable photonics, neuromorphic computing, and analog AI~\cite{shastri2021photonics}. Furthermore, our Ising machine is fully programmable, highly stable, and scalable---a combination of attributes typically compromised in previously demonstrated architectures.

Our work represents a substantial advancement in heuristic solvers for challenging combinatorial optimization problems, with meaningful implications for real-world applications such as protein folding~\cite{dill2012protein} and number partitioning problems~\cite{lucas2014ising}. These problems feature frustrated energy landscapes with a combinatorial explosion of metastable states, or local minima~\cite{nymeyer1998}, which pose a serious challenge to conventional local search heuristics. One of the central challenges for Ising machines is their limited ability to navigate frustrated landscapes, where numerous local extrema impede convergence to global optima--- a behavior characteristic of glassy systems. Whether governed by physical or simulated annealing, these problems remain inherently difficult to solve.  We chose protein folding and number partitioning as benchmarks because they typify glassy complexity, making them stringent tests for any Ising machine. While other photonic approaches---such as those based on spatial light modulators (SLMs)~\cite{honjo2021100} or degenerate optical parametric oscillators (DOPOs)~\cite{pierangeli2019large}---have demonstrated systems with large spin counts, none have addressed highly frustrated problems with ground-state accuracy.

\begin{landscape}
\begin{table}[ht]
\centering
\renewcommand{\arraystretch}{1.5}
\setlength{\tabcolsep}{1.5pt}

    \begin{tabular}{cccccccccccc} \hline \hline
 \multirow{2}{*}{Type}  & & Spin-spin  & Nonlinearity/&  \multirow{2}{*}{Frequency} & \multirow{2}{*}{Task performed} & Number of & \multirow{2}{*}{Connectivity} & Solution   & Number of &  \multirow{2}{*}{TTS} & \multirow{2}{*}{Year~[Ref]}  \vspace{-5pt}\\ 
   &   & interaction & spin update & & &spins $(N)$ & &quality& iterations&  \\ \hline
  \multirow{14}{*}{\textbf{OEO}}  &    & & & &\textbf{Number partitioning}  &  \textbf{256} &  \textbf{all-to-all} &   \textbf{100\%} &   \textbf{572} &  \textbf{14.08 s$^\dagger$ (353.64 $\mu$s$^\ast$)} \\ 

      & &&& &\textbf{Protein folding: S$_{30}$} & \textbf{630} &  \textbf{sparse} &  \textbf{99.7\%} &  \textbf{200} &  \textbf{2.34 s$^\dagger$ (100 $\mu$s$^\ast)$ }  \\
 &  &  \textbf{Optical} & \textbf{E/O} & \textbf{106 GBaud}& \textbf{Max-Cut: G22 graph}  & \textbf{2,000} &  \textbf{sparse}& \textbf{99.48\%} &  \textbf{800} & \textbf{12.30~s$^\dagger$ (400 $\mu$s$^\ast$)}  & \textbf{2025} \\

   &   &  & & & \textbf{Max-Cut: G81 graph} &\textbf{20,000} & \textbf{sparse}& \textbf{96.36\%} & \textbf{800} & \textbf{29.30 s$^\dagger$ (754.72 $\mu$s$^\ast$)} & \textbf{[This Work]} \\
         &  &  & & & \textbf{Square lattice}  & \textbf{41,209} & \textbf{sparse} & \textbf{97.3\%} & \textbf{1,000} & \textbf{75.16 s$^\dagger$ (1.936 ms$^\ast)$}  \\
    
    &   &  Optical & Electronic ASIC &1 GHz & 2-coloring graph & 64 & sparse & 100\% & 537 & 2.7 $\mu$s & 2025~\cite{hua2025} \\     
          &  & FPGA & E/O & 2.65 GBaud &Max-Cut: G22 graph & 2,000 & sparse & $\sim$97\% & 2,000 & $3.56~\text{ms}^\dagger$ &2024~\cite{li2024scalable} \\
   &   & Optical &  E/O & 6.25 GBaud&Square lattice & 16,384 &sparse & N/A &2,000 & $800.74~\text{s}^\dagger$ & 2024~\cite{li2024scalable} \\ 
     &   & Optical& E/O  & 0.25 GHz & Square lattice &  25,600 &sparse & 100\% & N/A & 22.4 $\mu$s  & 2022~\cite{cen2022large}\\ 
&    &&& & Möbius ladder  & 56 & sparse& 100\% & 20 & 4.5 $\mu$s \\ 
    &     & CPU & E/O  &  1 Hz & Square lattice & 100 &sparse & 100\% & 100 & 100 s  & 2019~\cite{bohm2019poor}\\
 \hline      
    &    & Optical & Optical & 1 GHz &  1D ring & 16 &sparse &  100 \% & 10,000 & N/A  & 2016~\cite{takata201616}\\

 \multirow{6}{*}{OPO} & & FPGA & Optical & 0.1 GHz & Möbius ladder & 100 & sparse & 100\% & 300 & 480 $\mu$s~ & 2016~\cite{mcmahon2016fully}\\ 
     
    &  &  &  & & Max-Cut: random cubic graph & 100 & all-to-all & 100\% & $<125,000^\dagger$ & $<$ 200 ms~ & \\ 

    &  &FPGA & Optical & 1 GHz & Max-Cut: K2000 graph & 2,000 & all-to-all & N/A & 1,000 &  5 ms~ & 2016~\cite{inagaki2016coherent}\\ 
    &  &  & & & Max-Cut: G22 graph & 2,000 & sparse & 99.17\% & 1,000 & 5 ms~ & \\ 
    &  & & & & Max-Cut: G39 graph & 2,000 & sparse & 96.68\% & 1,000 & 5 ms~ & \\

     &   & FPGA & Optical & 5 GHz & Max-Cut: random graph & 100,000 & all-to-all & N/A & 890 & 21.9~ms & 2021~\cite{honjo2021100}\\ 
 \hline 
 \multirow{6}{*}{ SLM}  & & Optical & Optical & N/A &  1D ring  & 7 & sparse  &  100\% & N/A & 150 ms  &2019~\cite{babaeian2019single} \\
 
  & & Optical& CPU & N/A &  Möbius ladder & 75,076 &  sparse & N/A & $\sim$1,000 & N/A & 2019~\cite{pierangeli2019large}  \\ 
&   & Optical &CPU& 100~Hz$^\dagger$ &  Möbius ladder & 16 & sparse  &  100\% & 600 & $6~s^\dagger$ &2023~\cite{WDM_Li2023} \\
  &   &Optical &CPU & N/A &  Max-Cut: random graph & 20,736 & all-to-all & N/A & 100 & 325 s & 2023~\cite{ye2023photonic}  \\ 
&    &Optical & CPU& N/A & Max-Cut: random graph & 30 & all-to-all  & 85 \% & 2000 & $640 ~s^\dagger$ &2024~\cite{ouyang2024demand} \\
    
\hline \hline      
\end{tabular}
%  \vspace{5pt}
\caption{\textbf{Survey of recent photonic Ising machine literature.} Each work is classified into one of three architecture categories: OEO (optoelectronic oscillator), OPO (optical parametric oscillator), or SLM (spatial light modulator). The ``Frequency'' column compares the intrinsic speed associated with each spin; for OEO architectures, this is a modulator baud rate or electronic clock rate; for OPO architectures, this is a pulse repetition rate; for SLM architectures, this is often not reported and difficult to estimate. E/O (optoelectronic).
$^\dagger$estimated; $^\ast$estimated with pipelined DSP; N/A not possible to estimate from manuscript.
}
\label{comparison-table}

\end{table}
\end{landscape}

Moreover, bit resolution constraints (for spin coupling strengths) in many hardware platforms limit the range of problems that can be encoded. In this work, we employ fully connected Ising spin configurations to partition a set of 256 numbers, achieving solution accuracy unmatched by previous photonic Ising machines. We also solve HP lattice protein folding problems, achieving parity with D-Wave's quantum system~\cite{irback2022}. Our photonic Ising machine folds protein chains of up to 30 amino acids ($S_{30}$) with 99\% of the ground state energy, and achieves 100\% accuracy for chains up to $S_{10}$. In contrast to D-Wave's quantum annealer, which operates at cryogenic temperatures and scales quadratically $\mathcal{O}(L^2)$ with amino acid sequence length, our room temperature photonic system scales linearly $\mathcal{O}(L)$ and requires no auxiliary spins. This work marks a significant step toward the realization of a large-scale, fully integrated photonic Ising machine capable of addressing a wide range of real-world optimization tasks, including drug discovery, material synthesis, and neural network optimization.

\section{Architecture and operational principle}
\label{sec:architecture}

Minimizing the energy function of the Ising model $H = \sum_{i < j} J_{ij} \sigma_i \sigma_j + \sum_i h_i \sigma_i$, with local fields $h_i$, couplings $J_{ij}$, and spin orientations $\sigma_i\in\pm1$~\cite{ising1924beitrag}, can be reformulated as a matrix-vector multiplication (MVM) problem. This is expressed using a weight matrix $\mathbf{W}$ and a spin vector $\mathbf{X}$, with $\mathbf{W}=[\alpha \mathbf{I} - \beta \mathbf{J} | -\beta\mathbf{h}]$, where $\alpha$ and $\beta$ represent feedback and coupling strengths, respectively~\cite{bohm2019poor}. The mapping between this weight matrix and the Ising Hamiltonian is detailed in the Methods, and its relationship to the Hopfield-inspired update equation is discussed in Supplementary Section~S1.4. Experimentally, we implement the Ising machine using two cascaded TFLN Mach-Zehnder modulators (MZMs)~\cite{lin2024} operating at 1310~nm in a closed-loop feedback configuration (Fig.~\ref{fig:cmim}). The first MZM encodes the time-multiplexed spins $\sigma_N$ as $\mathbf{X}(t)$, exploiting its nonlinear transfer function to approximate spin states. The second MZM linearly modulates the fixed interaction matrix $\mathbf{W}$. Optical element-wise multiplication is followed by electronic summation, facilitated by a QD SOA and photodetector (PD). High-speed electro-optic feedback is completed via ADC/DAC modules operating at 256~GSa/s, enabling iterative convergence toward the ground state. This OEO-based Ising machine is equivalent to a continuous-domain Hopfield neural network with controllable stochastic perturbations, and it heuristically solves complex optimization problems to yield high-quality solutions. Optimal hyperparameters ($\alpha,\beta$) are tuned per problem to ensure rapid convergence and stability. While ultrafast operation above 64~GBaud introduces signal impairments that can limit bit resolution and solution fidelity, we address these challenges using optimized DSP and operate successfully up to 128 GBaud~\cite{estaran2019}. Additionally, annealing is emulated via system-inherent electrical noise and a dedicated optical noise source through an SOA, accelerating convergence. Further experimental details are provided in the Methods.

\section{Results}
\label{sec:results}

\subsection{Bifurcation and spin-spin coupling at high speed}
\begin{figure}[h]
\centering
\includegraphics[width=1.0\textwidth]{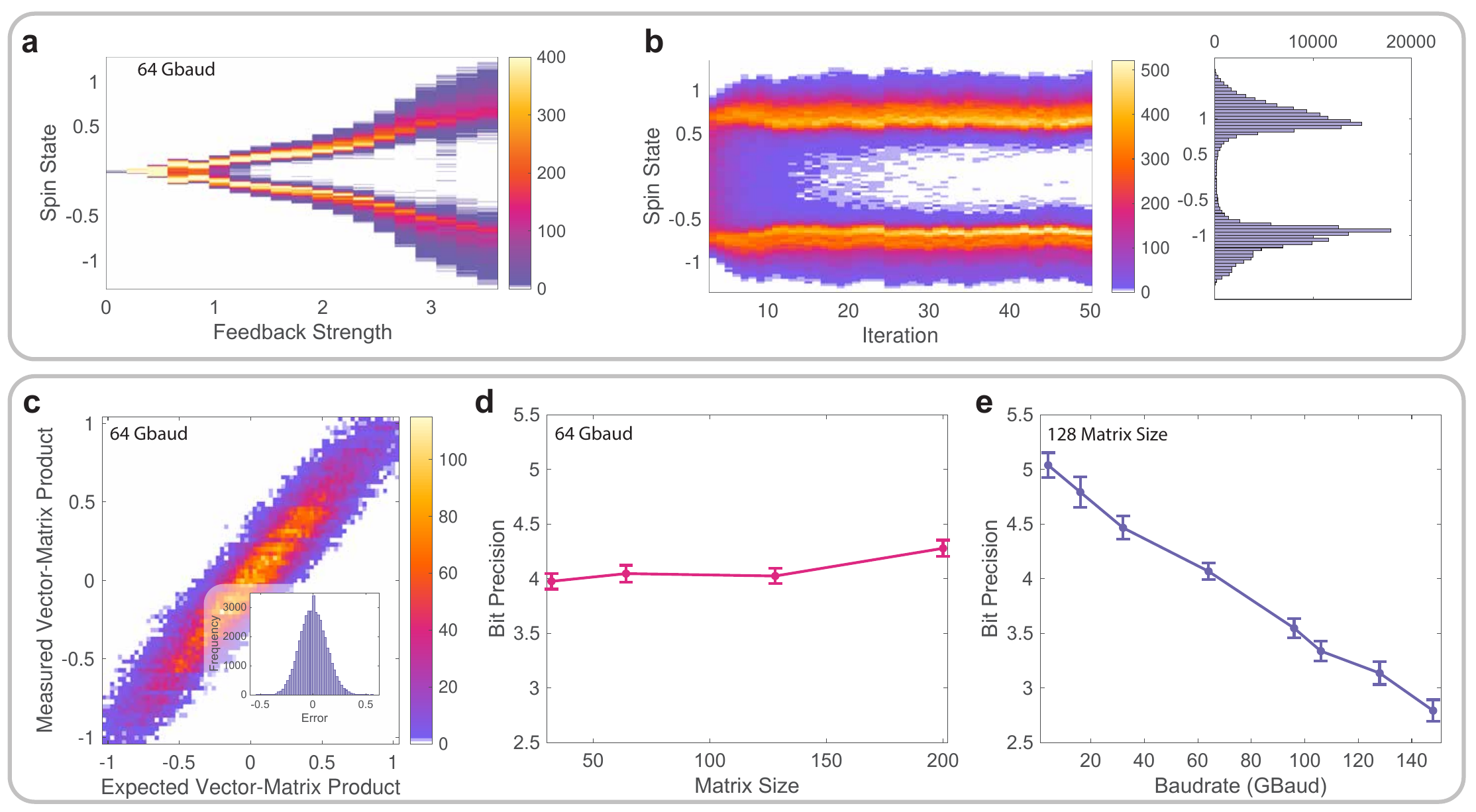}
 \caption{\textbf{Experimental characterization of bifurcation and matrix-vector multiplication.}
(a) Heatmap of the bifurcation dynamics as a function of feedback strength while operating at 64~GBaud with $N=$~262,144 uncoupled spins. The colour indicates the number of spins in a given state after 50 iterations. (b) Heatmap showing the evolution of uncoupled spins over successive iterations at a feedback strength of $\alpha=3.5$ (left), alongside a histogram of spin values at the final iteration (right). (c) Experimental results of matrix multiplication performed at 64~GBaud using 500 randomly sampled 128$\times$128 matrices. (d) Bit precision of matrix multiplication as a function of matrix size, ranging from $N=32$ to $N=200$. (e) As the symbol rate increases from 4~GBaud to 148~GBaud, the bit precision of matrix multiplication decreases linearly from 5.0342 to 2.7885.
}
\label{fig:bifurcation}
\end{figure}

Bifurcation is a fundamental feature of the Ising model, associated with phase transitions, and must be observable in any functional Ising machine~\cite{wang2023}. Its presence confirms that the system possesses sufficient nonlinearity and feedback strength to support the required dynamical behavior. To investigate this, we examined the nonlinear dynamics of the CMIM by characterizing bifurcation as a function of the feedback strength, $\alpha$. According to linear stability analysis, each spin node undergoes a transition from monostability to bistability when $\alpha$ exceeds a critical value $\alpha_0$~\cite{bohm2019poor}. To experimentally demonstrate bifurcation, we randomly initialized an ensemble of $N$ spin states operating at 64~GBaud and measured their amplitude while varying $\alpha$, in the absence of coupling ($\mathbf{J} = 0$). As shown in Fig.~\ref{fig:bifurcation}(a), all spins in the system initially converged to a single stable fixed point for $\alpha<\alpha_0$. When the feedback strength exceeded $\alpha>\alpha_0$, the system bifurcated into two stable fixed points. Fig.~\ref{fig:bifurcation}(b) illustrates the evolution of individual spins over 50 iterations at $\alpha=3.5$. Depending on their random initial amplitudes, each spin converged to one of the two stable fixed points, as shown in the resulting histogram. We observed consistent bifurcation behavior across baud rates from 64 to 128~GBaud. At lower baud rates ($<64$~GBaud), the DSP was able to compensate for frequency response distortions, enabling full bifurcation. At higher baud rates, however, accumulated errors became harder to correct, leading more spin states to settle between fixed points. Introducing a nonzero coupling matrix improved convergence by guiding spins toward stable configurations.

To further validate the system’s capability for Ising computation, we performed high-speed, large-scale MVM to demonstrate the required spin–spin coupling. Both TFLN MZMs were operated in the linear regime to assess MVM accuracy. The modulators were driven with two random vectors, $\mathbf{x}_1$ and $\mathbf{x}_2$, sampled from a uniform distribution in $[-1, 1]$ at 8-bit resolution. The feedforward result is an element-wise multiplication, $\mathbf{x}_1 \odot \mathbf{x}_2$, followed by digital summation to implement the multiply-and-accumulate (MAC) operation. At 106~GBaud, we achieved 212~GOPS on a single wavelength channel, nearly doubling the performance reported in prior work~\cite{lin2024}.  Fig.~\ref{fig:bifurcation}(c) compares measured and expected dot product values at 64~GBaud. From the measurements histogram, we estimate the accuracy to be $96.2\pm0.4$~\%. Fig.~\ref{fig:bifurcation}(d) shows this accuracy remained stable as the matrix size increased from 32 to 200. However, as shown in Fig.~\ref{fig:bifurcation}(e), increasing the symbol rate from 4~GBaud to 148~GBaud reduced accuracy from $98.16 \pm 0.31$~\% to $90.7 \pm 0.94$~\%. Correspondingly, the effective bit precision (calculated via adopting the method described in \cite{zhang2022silicon}) dropped from 4.5~bits at 32~GBaud to 3.3~bits at 106~GBaud. This reduction in analog MVM accuracy with increasing baud rate is primarily due to system noise and distortion, which scale with bandwidth. However, in the context of Ising machines, such degradation can be beneficial---acting as a form of annealing that helps the system escape local minima and converge toward lower-energy configurations~\cite{bohm2022noise}.

\subsection{Programmable Ising solver}
Having characterized the system's performance, we employed the CMIM to solve a range of benchmark problems. Its reconfigurability---including its support for arbitrary real-valued weights and flexible connectivity, from sparse to fully connected--- makes it a versatile heuristic solver for problem classes that are often difficult to implement on other architectures. We leveraged these capabilities to address large-scale tasks at ultrafast baud rates exceeding 100~GBaud.

\subsubsection{Benchmarking: square lattice and Max-Cut of Gset graphs}
\label{sec:benchmarking_maxcut}

\begin{figure}[ht]
\centering
\includegraphics[width=1.0\textwidth]{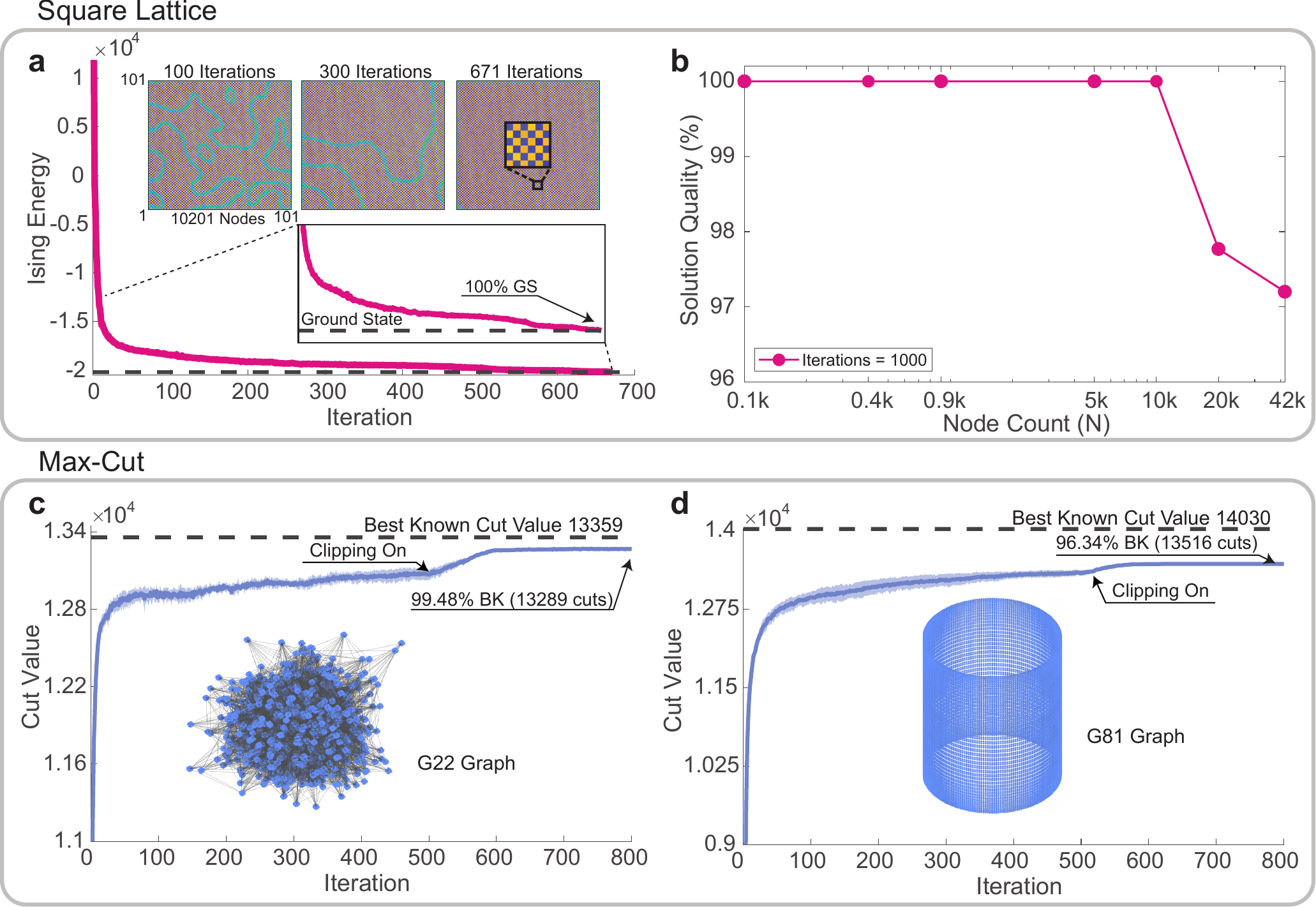}
\caption{\textbf{Benchmarking results.} Experimental results for the square lattice, G22 graph, and G81 graph problems, all evaluated at the maximum operating baud rate of 106~GBaud. (a) Time evolution of the Ising energy during ground state search for a 2D square lattice with 10,201 spins ($101\times 101$ lattice). The inset shows snapshots of the spin domain evolution across iterations, illustrating convergence towards the ground state solution. (b) Scaling analysis of solution quality--- measured by proximity to the ground state---across problem sizes ranging from 100 to 41,209 spins ($203\times 203$ lattice length). (c) Time evolution of the cut value for the G22 graph. The system achieves 99.48\% of the best-known cut, reaching a cut value of 13,289. The inset shows the corresponding coupling matrix with 1\% edge density and binary weights of 0 and 1. (d) Time evolution of the cut value for the G81 graph. The system reaches a solution quality exceeding 96.34\%, with a cut value of 13,516. The inset shows the G81 planar graph with weight values of $+1$ and $-1$. The evolution of cut values for both the G22 and G81 graphs is based on a fixed iteration count of 800.}
\label{fig:benchmarking}
\end{figure}

Two widely used benchmark problems for Ising machines are the 2D square lattice graph, which has a known polynomial-time solution, and the Max-Cut problem on Gset graphs, which is NP-hard in the general case~\cite{bohm2019poor}. Both can be formulated as Ising problems by appropriately configuring the weight matrix such that the minimum energy states correspond to optimal solutions (see Methods for mapping details). Fig.~\ref{fig:benchmarking}(a) shows experimental results for the 2D square lattice. The ground state corresponds to alternating neighboring spins, forming a checkerboard pattern. For each random initial spin configuration ($\boldsymbol\sigma$), the fixed weight matrix ($\mathbf{W}$) was defined by assigning antiferromagnetic coupling ($J_{ij} = -1$) to each lattice edge. For these experiments, hyperparameters $\alpha = 0.86$ and $\beta = 1$ yielded good convergence. As shown in the inset of Fig.~\ref{fig:benchmarking}(a), the system initially forms small spin domains that coalesce into a checkerboard pattern as the energy is minimized. For a $101\times 101$ lattice (10,201 spins with 50,601 edges), operating at 106~GBaud, the ground state was reached in $\sim$671 iterations (rightmost inset of Fig.~\ref{fig:benchmarking}(a)). The experiments were repeated 10 times to estimate the distribution of solution energies. To demonstrate scalability and reconfigurability, we varied the graph sizes from 100 nodes ($10 \times 10$ lattice) to 41,209 nodes ($203 \times 203$ lattice), as shown in Fig.~\ref{fig:benchmarking}(b). All experiments were capped at 1,000 iterations for consistent comparison. The system successfully reached the ground state for all lattices up to 10,201 nodes. For larger graphs, solution quality gradually decreased, achieving $\sim$97.2\% of the ground state energy at 41,209 nodes. To investigate this, in Supplementary Section S1.7.1, we present a simulated model of our experimental system, which shows that increasing the number of iterations and tuning hyperparameters enables convergence to the ground state. The Max-Cut problem aims to partition the vertices of a weighted graph into two subsets to maximize the number of edges crossing the cut. The Gset graphs are a collection of large, sparse, weighted and unweighted graphs with best-known cut values reported in the literature~\cite{benlic2013}. Our system achieved competitive cut values across multiple Gset graphs. For the G22 graph (2,000 vertices and 19,990 edges, with 41,980 nonzero elements), we achieved 99.48\% of the best-known value of 13,352 (see Fig. 3(c)). For the G81 graph (20,000 spin nodes and 40,000 edges, corresponding to 100,000 nonzero elements), the system also demonstrated fast convergence (Fig. 3(d)). The experimentally achieved cut values for both graphs surpass those reported for all prior photonic Ising machines~\cite{gao2024photonic} (see Supplementary Table~S4). This performance gain is primarily attributed to: (1) the controllable optical noise source (bulk SOA), combined with exponential annealing to escape local minima, described in Method Section~\ref{sec:optical_noise}; and (2) signal clipping during DSP, which mitigates amplitude inhomogeneity. Further improvements are expected with advanced algorithms, including chaotic amplitude control~\cite{leleu2021}, momentum-based methods~\cite{kalinin2023analog}, or variable step-size techniques~\cite{pramanik2024}. To compare our hardware to a software-based computer simulation, in Supplementary Section S3.2, we limited the bit depth of our simulated model and ran it on the G22 graph. Notably, the simulation required 6-bit precision to match the solution quality of our analog hardware, which achieved similar performance at 3.3-bit effective resolution while operating at 106~GBaud. This confirms that comparable or superior solution quality can be attained at lower bit precision. The interplay of continuous optical waveforms, nonlinear modulation, and modest DSP overhead enabled our system to avoid the computational cost typically associated with high-precision digital computation.

\subsubsection{Folding lattice proteins} 
\begin{figure}[ht]
\centering
\includegraphics[width=1.0\textwidth]{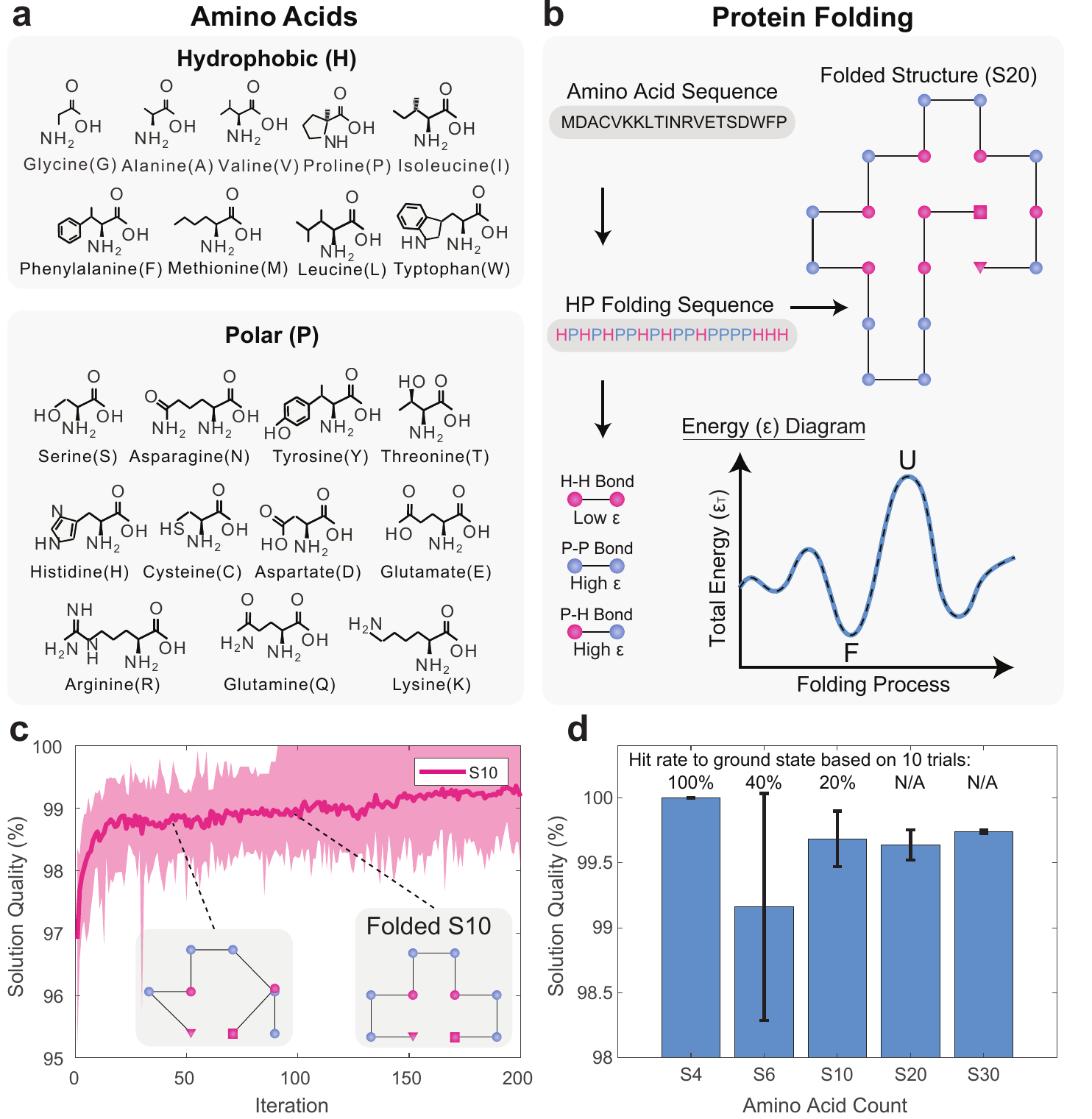}
 \caption{\textbf{HP model protein folding problem.} (a) The 20 naturally occurring amino acids are categorized as hydrophobic (H) and polar (P). (b) Mapping of an amino acid sequence onto the HP lattice, where only H--H interactions contribute to lowering the energy. The native (folded) configuration maximizes H--H interactions. The sequence transitions from an unfolded (U) to a folded (F) configuration. The ground state energy for a given sequence corresponds to the solution of the lattice folding problem. (c) Evolution of the energy landscape over iterations for an amino acid sequence length $S_{10}$. The inset illustrates misfolded and correctly folded configurations at two iterations. (d) Experimentally obtained solution quality for amino acid sequences ranging from $S_{4}$ to $S_{30}$.}
\label{fig:protein_folding}
\end{figure}

The goal of computational protein folding is to predict the stable physical structure a sequence of amino acids will adopt. This remains a major challenge due to the exponentially growing solution space~\cite{dill2012protein}. Recent deep learning models such as AlphaFold 3 have dramatically improved our ability to predict complex protein structures~\cite{abramson2024}. However, energy-based models, such as the simplified HP lattice model, offer a complementary approach by directly encoding the physics of amino acid interactions, providing mechanistic insights into folding dynamics. In the HP model, each amino acid is classified as either hydrophobic (H) and polar (P), rather than one of the 20 natural amino acids (see Fig.~\ref{fig:protein_folding}(a)), and the lowest energy configuration maximizes the number of H--H interactions~\cite{irback2022}. Protein folding is thus framed as an optimization problem, where the native structure corresponds to the minimum-energy configuration (see Fig.~\ref{fig:protein_folding}(b)). In our experiment, we examined five standard 2D HP sequences of increasing lengths ($S_4, S_6, S_{10}, S_{20}, S_{30}$). The Methods describes how the folding process for a sequence was mapped to an equivalent Ising formulation. Fig.~\ref{fig:benchmarking}(c) shows the solution quality evolution for the $S_{10}$ sequence. The ground-state configuration was achieved within 100 iterations, with an average solution quality of 99\% across 10 runs. The inset shows representative misfolded and correctly folded structures over the course of optimization. Fig.~\ref{fig:benchmarking}(d) summarizes the average solution quality and hit rate (the proportion of trials that reached the ground state) for all five sequences. We observed a 100\% hit rate for $S_{4}$, which decreased with sequence length. Although the ground state was not achieved for sequences longer than $S_{10}$, our hardware consistently maintained high solution quality, exceeding 99\% for $S_{30}$. Hit rates are expected to improve with increased iterations and integration of advanced algorithms, such as chaotic amplitude control~\cite{leleu2021} or momentum-based techniques~\cite{kalinin2023analog}, particularly for longer sequences. In principle, our approach can be extended to more complex protein folding models, including 3D lattices or the Miyazawa-Jernigan model, which incorporates all 20 amino acid types. Our results are comparable to those achieved using quantum annealing~\cite{irback2022}, but with significant architectural advantages. Unlike D-Wave's quantum annealer, which imposes a fixed qubit connectivity and requires graph embedding, the photonic CMIM natively supports arbitrary all-to-all spin connectivity. Embedding dense graphs on limited-connectivity hardware requires auxiliary spin variables, introducing a quadratic overhead: $N_p\sim\mathcal{O}(N^2)$, where $N_p$ is the number of physical qubits required to encode a graph with $N$ logical nodes (see Supplementary Section S3.1). Moreover, D-Wave's performance is constrained by its limited physical qubit count, which restricts the maximum sequence length it can handle. In contrast, our photonic Ising machine is not subject to such limitations, offering a scalable path toward solving larger and more complex folding problems.

\subsubsection{Number partitioning}

\begin{figure}[h]
\centering\includegraphics[width=1.0\textwidth]{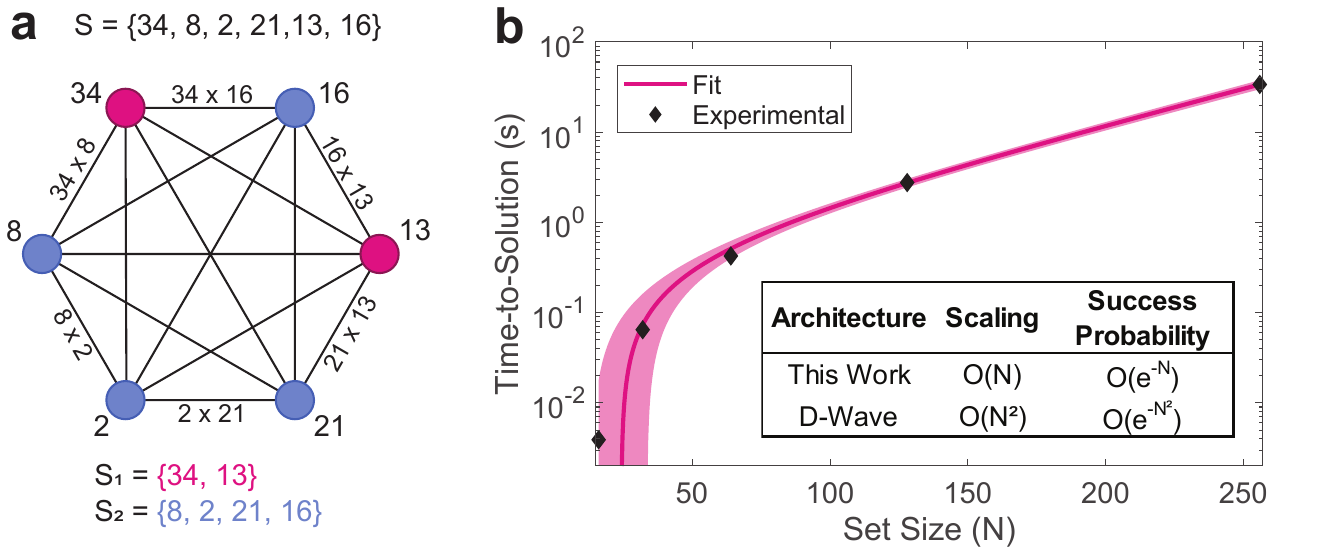}
\caption{\textbf{Number partitioning problem.} (a) Conceptual illustration of the number partitioning problem, modeled as an all-to-all connected graph, where each spin represents an assignment of a number to one of two equal-sum subsets. (b) Computation time to reach the ground state exhibits exponential scaling with problem size, following the model $Ae^{bN}+C$, where $A=0.300057$, $b=0.018509$, $C=-0.469471$. Ground state solutions were obtained at 106~GBaud. The inset table summarizes the required number of spins with problem size and the success probabilities for both the CMIM and D-Wave's quantum system~\cite{hamerly2019experimental}.}
\label{fig:number_partitioning}
\end{figure}

Number partitioning---dividing a set of integers into two subsets with equal sums---is a well-known combinatorial optimization problem with an exponentially growing solution space of $\mathcal{O}(2^{N})$~\cite{lucas2014ising}. Despite significant interest, photonic hardware has yet to demonstrate large-scale, high-speed, and high-fidelity performance on this task~\cite{huang2021,prabhakar2023,asproni2020accuracy, xu2020}. Fig.~\ref{fig:number_partitioning}(a) illustrates a schematic of the number partitioning problem for a six-node, fully connected Ising system, where coupling strengths are determined by the values of the set elements (see Methods). Starting from a randomly initialized $N$-spin configuration, the system converges into two spin subsets that encode the optimal partition. In our experiment, we successfully partitioned sets with problem sizes ranging from 16 to 256, achieving ground-state solutions. Fig.~\ref{fig:number_partitioning}(b) shows the experimentally obtained average TTS for different problem sizes calculated using the method described in Supplementary Section S1.3. A curve fit indicates that TTS increases exponentially with problem size, consistent with the complexity of the task. In several cases, the system returned multiple valid solutions within a fixed 1,000-iteration window, likely due to the inherent system noise facilitating exploration of degenerate ground states (see Supplementary Section S1.7.2).
To compare success probabilities with D-Wave's quantum processing unit (QPU), we mapped the number partitioning problem onto its architecture (see Methods) and conducted experiments with annealing times ranging from 0.5 to 500 $\mu$s. Fig.~S14 shows that the QPU achieved a success probability of 83.9\% at $N=4$, which dropped to 0.03\% for $N=32$. Even after $10^4$ runs at the maximum annealing time of 500~$\mu$s, the QPU failed to reach the ground state for $N > 32$. This limitation arises from the QPU's graph embedding requirement, which imposes again a quadratic scaling overhead in physical qubits, resulting in a success probability that scales as $P \sim \mathcal{O}(e^{-N^2})$ (see Supplementary Section S3.1). In contrast, our photonic Ising machine achieved a 100\% success rate for number partitioning problems up to $N = 256$, with performance bounded only by internal memory limits in the current experimental setup that can be relaxed with enhanced hardware. To explore scalability beyond experimental limits, we simulated the CMIM setup for problems with sizes from $N=4$ to $N=1024$. The results show that our Ising machine maintained 100\% success probability up to $N = 285$. Beyond this, the success rate followed an exponential decay, $P \sim \mathcal{O}(e^{-N})$, reaching $\sim$0.15\% at $N = 1024$.  

\section{Discussion and conclusion}
\label{sec:discussion}
In addition to supporting reconfigurable weight matrices, large-scale spins, and high baud rates, our photonic architecture is robust against ambient variations such as voltage drift, optical power changes, and temperature fluctuations. It exhibits negligible variation in chromatic dispersion, loss, and nonlinear optical effects over the range $15^{\circ}\mathrm{C}$ to $45^{\circ}\mathrm{C}$ (see Supplementary Section S1.2). Whereas most Ising architectures are often highly sensitive to phase and amplitude variations~\cite{honjo2021100, cen2022large}, our platform can operate continuously without active stabilization. This capability for uninterrupted long-term operation is critical for addressing complex combinatorial problems, particularly at high speeds and large scales. Additionally, the system's DSP algorithm compensates for hardware imperfections such as nonuniform spin amplitudes and bandwidth limitations, enabling convergence to high-quality solutions. The DSP employed is an industry-standard linear processor, contributing minimally to overall energy consumption~\cite{tariq2020computational}. In the current setup, feedback latency is constrained by discrete DAC/ADC components; the limitations of the DSP are detailed in Supplementary Section S2. These limitations can be significantly reduced using a single-platform solution with pipelined DSP. Assuming such a DSP pipeline supports throughput at the modulator baud rate, the feedback latency is constrained only by the processing latency of a single symbol, and the total CMIM iteration latency is given by the greater of the feedforward and feedback latencies. For the algorithms employed in this work, we estimate a pipeline depth of 115, comprising a 20-tap anti-aliasing filter, a 51-tap feedforward equalization filter, 3 additional cycles for sample summation, and a 51-tap pulse-shaping filter. With a clock rate in the range 200--500~MHz, this configuration would yield an upper bound of around 500~ns feedback latency. As a result, the G22 graph problem, with a feedforward latency of $\sim$377 ns, would achieve total TTS of $\sim$400 $\mu$s for 800 iterations with such an implementation---more than 10$\times$ faster than the CIM \cite{inagaki2016} (see Supplementary Section S1.3). The required capabilities are supported by commercially available ASICs, such as those in the Ciena Wavelogic transceiver family~\cite{cienaWaveLogic}. For example, the Wavelogic 6 Extreme features ADCs/DACs with a 100~GHz 3~dB bandwidth and 225~GSa/s sampling rate, and embeds DSP capable of executing the algorithms used in this work, supporting up to 200~GBaud operation. Recent work~\cite{pappas2025,lin2024} has shown that a hypermultiplexed architecture, combining time-, space-, and wavelength-division multiplexing (TDM, SDM, WDM), can significantly enhance computational throughput~\cite{bai2023}. This multidimensional multiplexing strategy could enable a \textit{parallel CMIM} capable of reducing latency for a single trajectory of the Ising algorithm or to solve multiple independent trajectories in parallel. In Supplementary Section S4, we analyzed the compute throughput and energy efficiency of this parallel architecture. Our current implementation (single-wavelength, single-channel) achieves a compute efficiency of 48~GOPS/W. For comparison, state-of-the-art single instruction/multiple data (SIMD) architectures such as the NVIDIA H100 GPU deliver 95.6~GOPS/W at FP64 precision~\cite{nvidiaNVIDIAH100Tensor}. However, by leveraging commercially available technologies, a parallel CMIM with 16 channels could achieve a throughput of 868~TOPS with an efficiency of 2.34~TOPS/W, supporting over 10 million spins. With further advancements, the architecture could scale to over 30 million spins, delivering a throughput of 9.3~POPS at 5.9~TOPS/W efficiency, surpassing the H100's FP8 precision performance of 1.98~POPS at 2.83~TOPS/W~\cite{nvidiaNVIDIAH100Tensor}. \vspace{3pt}\\
We have demonstrated a large-scale, programmable, and highly stable CMIM, enabled by TFLN photonics within a Hopfield-network-inspired framework. This system supports up to 256 fully connected spins (or 41,209 sparsely connected spins) and operates at a baud rate of 106~GBaud. The integration of DSP techniques significantly enhances solution quality and convergence rate, particularly at high speeds and large scales. Our work demonstrates the largest spin configuration in any OEO-based Ising machine, and it achieves best-in-class performance on standard benchmark problems, outperforming previously reported photonic Ising machines. Compared to electronic platforms, our approach offers compelling advantages in energy efficiency and scalability \cite{gao2024photonic}. Additionally, while delivering comparable performance to D-Wave's quantum annealer on protein folding tasks, it significantly outperforms on number partitioning problems. Unlike quantum systems, our machine operates at room temperature, avoids topological embedding constraints, and exhibits linear scaling for dense problems versus quadratic scaling imposed by hardware-limited connectivity in quantum annealers. Note that there is a large class of problems addressed in the literature that employ sparsely connected graphs (see the discussion of application sparsity in Supplementary Section S2). In general, solving combinatorial optimization problems using quantum hardware faces several critical challenges, including: (1) Maintaining quantum coherence, which is highly susceptible to noise and environmental disturbances; (2) Achieving scalable qubit integration for practical quantum algorithms; and (3) Developing robust quantum error correction schemes to mitigate high intrinsic error rates associated with quantum operations~\cite{gill2024quantum, de2021materials}. In contrast, our work marks a significant step forward in realizing a scalable, fully-integrated, high-speed photonic Ising machine, with broad applications in drug discovery, material synthesis, prime factorization, and machine learning.

%\section*{Online content}
%Any methods, additional references, Nature Portfolio reporting summaries, extended data, supplementary information, acknowledgements, peer review information, details of author contributions and competing interests, and statements of data and code availability are available at https://doi.org/xxx/xxx

%\bibliography{sn-bibliography}

\begin{thebibliography}{50}
\ifx \bisbn   \undefined \def \bisbn  #1{ISBN #1}\fi
\ifx \binits  \undefined \def \binits#1{#1}\fi
\ifx \bauthor  \undefined \def \bauthor#1{#1}\fi
\ifx \batitle  \undefined \def \batitle#1{#1}\fi
\ifx \bjtitle  \undefined \def \bjtitle#1{#1}\fi
\ifx \bvolume  \undefined \def \bvolume#1{\textbf{#1}}\fi
\ifx \byear  \undefined \def \byear#1{#1}\fi
\ifx \bissue  \undefined \def \bissue#1{#1}\fi
\ifx \bfpage  \undefined \def \bfpage#1{#1}\fi
\ifx \blpage  \undefined \def \blpage #1{#1}\fi
\ifx \burl  \undefined \def \burl#1{\textsf{#1}}\fi
\ifx \doiurl  \undefined \def \doiurl#1{\url{https://doi.org/#1}}\fi
\ifx \betal  \undefined \def \betal{\textit{et al.}}\fi
\ifx \binstitute  \undefined \def \binstitute#1{#1}\fi
\ifx \binstitutionaled  \undefined \def \binstitutionaled#1{#1}\fi
\ifx \bctitle  \undefined \def \bctitle#1{#1}\fi
\ifx \beditor  \undefined \def \beditor#1{#1}\fi
\ifx \bpublisher  \undefined \def \bpublisher#1{#1}\fi
\ifx \bbtitle  \undefined \def \bbtitle#1{#1}\fi
\ifx \bedition  \undefined \def \bedition#1{#1}\fi
\ifx \bseriesno  \undefined \def \bseriesno#1{#1}\fi
\ifx \blocation  \undefined \def \blocation#1{#1}\fi
\ifx \bsertitle  \undefined \def \bsertitle#1{#1}\fi
\ifx \bsnm \undefined \def \bsnm#1{#1}\fi
\ifx \bsuffix \undefined \def \bsuffix#1{#1}\fi
\ifx \bparticle \undefined \def \bparticle#1{#1}\fi
\ifx \barticle \undefined \def \barticle#1{#1}\fi
\bibcommenthead
\ifx \bconfdate \undefined \def \bconfdate #1{#1}\fi
\ifx \botherref \undefined \def \botherref #1{#1}\fi
%\ifx \url \undefined \def \url#1{\textsf{#1}}\fi
\ifx \bchapter \undefined \def \bchapter#1{#1}\fi
\ifx \bbook \undefined \def \bbook#1{#1}\fi
\ifx \bcomment \undefined \def \bcomment#1{#1}\fi
\ifx \oauthor \undefined \def \oauthor#1{#1}\fi
\ifx \citeauthoryear \undefined \def \citeauthoryear#1{#1}\fi
\ifx \endbibitem  \undefined \def \endbibitem {}\fi
\ifx \bconflocation  \undefined \def \bconflocation#1{#1}\fi
\ifx \arxivurl  \undefined \def \arxivurl#1{\textsf{#1}}\fi
\csname PreBibitemsHook\endcsname

%%% 1
\bibitem[\protect\citeauthoryear{Mohseni et~al.}{2022}]{mohseni2022ising}
\begin{barticle}
\bauthor{\bsnm{Mohseni}, \binits{N.}},
\bauthor{\bsnm{McMahon}, \binits{P.L.}},
\bauthor{\bsnm{Byrnes}, \binits{T.}}:
\batitle{{Ising} machines as hardware solvers of combinatorial optimization problems}.
\bjtitle{Nature Reviews Physics}
\bvolume{4},
\bfpage{363}--\blpage{379}
(\byear{2022})
\end{barticle}
\endbibitem

%%% 2
\bibitem[\protect\citeauthoryear{Hamerly et~al.}{2019}]{hamerly2019experimental}
\begin{barticle}
\bauthor{\bsnm{Hamerly}, \binits{R.}},
\bauthor{\bsnm{Inagaki}, \binits{T.}},
\bauthor{\bsnm{McMahon}, \binits{P.L.}},
\bauthor{\bsnm{Venturelli}, \binits{D.}},
\bauthor{\bsnm{Marandi}, \binits{A.}},
\bauthor{\bsnm{Onodera}, \binits{T.}},
\bauthor{\bsnm{Ng}, \binits{E.}},
\bauthor{\bsnm{Langrock}, \binits{C.}},
\bauthor{\bsnm{Inaba}, \binits{K.}},
\bauthor{\bsnm{Honjo}, \binits{T.}}, \betal:
\batitle{Experimental investigation of performance differences between coherent {Ising} machines and a quantum annealer}.
\bjtitle{Science Advances}
\bvolume{5},
\bfpage{0823}
(\byear{2019})
\end{barticle}
\endbibitem

%%% 3
\bibitem[\protect\citeauthoryear{Hopfield}{1982}]{hopfield1982}
\begin{barticle}
\bauthor{\bsnm{Hopfield}, \binits{J.J.}}:
\batitle{Neural networks and physical systems with emergent collective computational abilities.}
\bjtitle{Proceedings of the National Academy of Sciences}
\bvolume{79},
\bfpage{2554}--\blpage{2558}
(\byear{1982})
\end{barticle}
\endbibitem

%%% 4
\bibitem[\protect\citeauthoryear{Lucas}{2014}]{lucas2014ising}
\begin{botherref}
\oauthor{\bsnm{Lucas}, \binits{A.}}:
{Ising} formulations of many {NP} problems.
Frontiers in Physics
\textbf{2},
\bfpage{5}
(2014)
\end{botherref}
\endbibitem

%%% 5
\bibitem[\protect\citeauthoryear{Dill and MacCallum}{2012}]{dill2012protein}
\begin{barticle}
\bauthor{\bsnm{Dill}, \binits{K.A.}},
\bauthor{\bsnm{MacCallum}, \binits{J.L.}}:
\batitle{The protein-folding problem, 50 years on}.
\bjtitle{Science}
\bvolume{338},
\bfpage{1042}--\blpage{1046}
(\byear{2012})
\end{barticle}
\endbibitem

%%% 6
\bibitem[\protect\citeauthoryear{Ising}{1925}]{ising1924beitrag}
\begin{barticle}
\bauthor{\bsnm{Ising}, \binits{E.}}:
\batitle{Beitrag zur theorie des ferromagnetismus}.
\bjtitle{Z. Physik}
\bvolume{31},
\bfpage{253}--\blpage{258}
(\byear{1925})
\end{barticle}
\endbibitem

%%% 7
\bibitem[\protect\citeauthoryear{Budrikis}{2024}]{budrikis}
\begin{barticle}
\bauthor{\bsnm{Budrikis}, \binits{Z.}}:
\batitle{100 years of the {Ising} model}.
\bjtitle{Nature Reviews Physics}
\bvolume{6},
\bfpage{530}
(\byear{2024})
\end{barticle}
\endbibitem

%%% 8
\bibitem[\protect\citeauthoryear{Garey and Johnson}{1979}]{garey1979}
\begin{bbook}
\bauthor{\bsnm{Garey}, \binits{M.R.}},
\bauthor{\bsnm{Johnson}, \binits{D.S.}}:
\bbtitle{Computers and Intractability: A Guide to the Theory of {NP}-completeness}.
\bsertitle{Mathematical Sciences Series},
\bvolume{174},
\bfpage{338}
(\byear{1979})
\end{bbook}
\endbibitem

%%% 9
\bibitem[\protect\citeauthoryear{McMahon}{2023}]{mcmahon2023physics}
\begin{barticle}
\bauthor{\bsnm{McMahon}, \binits{P.L.}}:
\batitle{The physics of optical computing}.
\bjtitle{Nature Reviews Physics}
\bvolume{5},
\bfpage{717}--\blpage{734}
(\byear{2023})
\end{barticle}
\endbibitem

%%% 10
\bibitem[\protect\citeauthoryear{Shastri et~al.}{2021}]{shastri2021photonics}
\begin{barticle}
\bauthor{\bsnm{Shastri}, \binits{B.J.}},
\bauthor{\bsnm{Tait}, \binits{A.N.}},
\bauthor{\bsnm{Lima}, \binits{T.}},
\bauthor{\bsnm{Pernice}, \binits{W.H.}},
\bauthor{\bsnm{Bhaskaran}, \binits{H.}},
\bauthor{\bsnm{Wright}, \binits{C.D.}},
\bauthor{\bsnm{Prucnal}, \binits{P.R.}}:
\batitle{Photonics for artificial intelligence and neuromorphic computing}.
\bjtitle{Nature Photonics}
\bvolume{15},
\bfpage{102}--\blpage{114}
(\byear{2021})
\end{barticle}
\endbibitem

%%% 11
\bibitem[\protect\citeauthoryear{Zhang et~al.}{2025}]{ZhangReview2025}
\begin{botherref}
\oauthor{\bsnm{Zhang}, \binits{T.}},
\oauthor{\bsnm{Liu}, \binits{S.}},
\oauthor{\bsnm{Jiang}, \binits{H.}},
\oauthor{\bsnm{Gross}, \binits{W.J.}},
\oauthor{\bsnm{Lombardi}, \binits{F.}},
\oauthor{\bsnm{Han}, \binits{J.}}:
Approximate and stochastic Ising machines.
IEEE Nanotechnology Magazine,
\bvolume{19},1--10
(2025)
\end{botherref}
\endbibitem

%%% 12
\bibitem[\protect\citeauthoryear{Gao et~al.}{2024}]{gao2024photonic}
\begin{botherref}
\oauthor{\bsnm{Gao}, \binits{Y.}},
\oauthor{\bsnm{Chen}, \binits{G.}},
\oauthor{\bsnm{Qi}, \binits{L.}},
\oauthor{\bsnm{Fu}, \binits{W.}},
\oauthor{\bsnm{Yuan}, \binits{Z.}},
\oauthor{\bsnm{Danner}, \binits{A.J.}}:
Photonic Ising machines for combinatorial optimization problems.
Applied Physics Reviews
\textbf{11}
(2024)
\end{botherref}
\endbibitem

%%% 13
\bibitem[\protect\citeauthoryear{Lin et~al.}{2024}]{lin2024}
\begin{barticle}
\bauthor{\bsnm{Lin}, \binits{Z.}},
\bauthor{\bsnm{Shastri}, \binits{B.J.}},
\bauthor{\bsnm{Yu}, \binits{S.}},
\bauthor{\bsnm{Song}, \binits{J.}},
\bauthor{\bsnm{Zhu}, \binits{Y.}},
\bauthor{\bsnm{Safarnejadian}, \binits{A.}},
\bauthor{\bsnm{Cai}, \binits{W.}},
\bauthor{\bsnm{Lin}, \binits{Y.}},
\bauthor{\bsnm{Ke}, \binits{W.}},
\bauthor{\bsnm{Hammood}, \binits{M.}},
\bauthor{\bsnm{Wang}, \binits{T.}},
\bauthor{\bsnm{Xu}, \binits{M.}},
\bauthor{\bsnm{Zheng}, \binits{Z.}},
\bauthor{\bsnm{Al-Qadasi}, \binits{M.}},
\bauthor{\bsnm{Esmaeeli}, \binits{O.}},
\bauthor{\bsnm{Rahim}, \binits{M.}},
\bauthor{\bsnm{Pakulski}, \binits{G.}},
\bauthor{\bsnm{Schmid}, \binits{J.}},
\bauthor{\bsnm{Barrios}, \binits{P.}},
\bauthor{\bsnm{Jiang}, \binits{W.}},
\bauthor{\bsnm{Morison}, \binits{H.}},
\bauthor{\bsnm{Mitchell}, \binits{M.}},
\bauthor{\bsnm{Guan}, \binits{X.}},
\bauthor{\bsnm{Jaeger}, \binits{N.A.F.}},
\bauthor{\bsnm{Rusch}, \binits{L.A.}},
\bauthor{\bsnm{Shekhar}, \binits{S.}},
\bauthor{\bsnm{Shi}, \binits{W.}},
\bauthor{\bsnm{Yu}, \binits{S.}},
\bauthor{\bsnm{Cai}, \binits{X.}},
\bauthor{\bsnm{Chrostowski}, \binits{L.}}:
\batitle{120 {GOPS} photonic tensor core in thin-film lithium niobate for inference and in situ training}.
\bjtitle{Nature Communications}
\bvolume{15},
\bfpage{9081}
(\byear{2024})
\end{barticle}
\endbibitem

%%% 14
\bibitem[\protect\citeauthoryear{Kalinin and Berloff}{2022}]{kalinin2022}
\begin{barticle}
\bauthor{\bsnm{Kalinin}, \binits{K.P.}},
\bauthor{\bsnm{Berloff}, \binits{N.G.}}:
\batitle{Computational complexity continuum within {Ising} formulation of {NP} problems}.
\bjtitle{Communications Physics}
\bvolume{5},
\bfpage{20}
(\byear{2022})
\end{barticle}
\endbibitem

%%% 15
\bibitem[\protect\citeauthoryear{Li et~al.}{2024}]{li2024scalable}
\begin{barticle}
\bauthor{\bsnm{Li}, \binits{Z.}},
\bauthor{\bsnm{Gan}, \binits{R.}},
\bauthor{\bsnm{Chen}, \binits{Z.}},
\bauthor{\bsnm{Deng}, \binits{Z.}},
\bauthor{\bsnm{Gao}, \binits{R.}},
\bauthor{\bsnm{Chen}, \binits{K.}},
\bauthor{\bsnm{Guo}, \binits{C.}},
\bauthor{\bsnm{Zhang}, \binits{Y.}},
\bauthor{\bsnm{Liu}, \binits{L.}},
\bauthor{\bsnm{Yu}, \binits{S.}}, \betal:
\batitle{Scalable on-chip optoelectronic {Ising} machine utilizing thin-film lithium niobate photonics}.
\bjtitle{ACS Photonics}
\bvolume{11},
\bfpage{1703}--\blpage{1714}
(\byear{2024})
\end{barticle}
\endbibitem

%%% 16
\bibitem[\protect\citeauthoryear{Nymeyer et~al.}{1998}]{nymeyer1998}
\begin{barticle}
\bauthor{\bsnm{Nymeyer}, \binits{H.}},
\bauthor{\bsnm{Garc{\'\i}a}, \binits{A.E.}},
\bauthor{\bsnm{Onuchic}, \binits{J.N.}}:
\batitle{Folding funnels and frustration in off-lattice minimalist protein landscapes}.
\bjtitle{Proceedings of the National Academy of Sciences}
\bvolume{95}(\bissue{11}),
\bfpage{5921}--\blpage{5928}
(\byear{1998})
\end{barticle}
\endbibitem

%%% 17
\bibitem[\protect\citeauthoryear{Honjo et~al.}{2021}]{honjo2021100}
\begin{barticle}
\bauthor{\bsnm{Honjo}, \binits{T.}},
\bauthor{\bsnm{Sonobe}, \binits{T.}},
\bauthor{\bsnm{Inaba}, \binits{K.}},
\bauthor{\bsnm{Inagaki}, \binits{T.}},
\bauthor{\bsnm{Ikuta}, \binits{T.}},
\bauthor{\bsnm{Yamada}, \binits{Y.}},
\bauthor{\bsnm{Kazama}, \binits{T.}},
\bauthor{\bsnm{Enbutsu}, \binits{K.}},
\bauthor{\bsnm{Umeki}, \binits{T.}},
\bauthor{\bsnm{Kasahara}, \binits{R.}}, \betal:
\batitle{100,000-spin coherent {Ising} machine}.
\bjtitle{Science Advances}
\bvolume{7},
\bfpage{0952}
(\byear{2021})
\end{barticle}
\endbibitem

%%% 18
\bibitem[\protect\citeauthoryear{Pierangeli et~al.}{2019}]{pierangeli2019large}
\begin{barticle}
\bauthor{\bsnm{Pierangeli}, \binits{D.}},
\bauthor{\bsnm{Marcucci}, \binits{G.}},
\bauthor{\bsnm{Conti}, \binits{C.}}:
\batitle{Large-scale photonic {Ising} machine by spatial light modulation}.
\bjtitle{Physical Review Letters}
\bvolume{122},
\bfpage{213902}
(\byear{2019})
\end{barticle}
\endbibitem

%%% 19
\bibitem[\protect\citeauthoryear{Hua et~al.}{2025}]{hua2025}
\begin{barticle}
\bauthor{\bsnm{Hua}, \binits{S.}},
\bauthor{\bsnm{Divita}, \binits{E.}},
\bauthor{\bsnm{Yu}, \binits{S.}},
\bauthor{\bsnm{Peng}, \binits{B.}},
\bauthor{\bsnm{Roques-Carmes}, \binits{C.}},
\bauthor{\bsnm{Su}, \binits{Z.}},
\bauthor{\bsnm{Chen}, \binits{Z.}},
\bauthor{\bsnm{Bai}, \binits{Y.}},
\bauthor{\bsnm{Zou}, \binits{J.}},
\bauthor{\bsnm{Zhu}, \binits{Y.}}, \betal:
\batitle{An integrated large-scale photonic accelerator with ultralow latency}.
\bjtitle{Nature}
\bvolume{640},
\bfpage{361}--\blpage{367}
(\byear{2025})
\end{barticle}
\endbibitem

%%% 20
\bibitem[\protect\citeauthoryear{Cen et~al.}{2022}]{cen2022large}
\begin{barticle}
\bauthor{\bsnm{Cen}, \binits{Q.}},
\bauthor{\bsnm{Ding}, \binits{H.}},
\bauthor{\bsnm{Hao}, \binits{T.}},
\bauthor{\bsnm{Guan}, \binits{S.}},
\bauthor{\bsnm{Qin}, \binits{Z.}},
\bauthor{\bsnm{Lyu}, \binits{J.}},
\bauthor{\bsnm{Li}, \binits{W.}},
\bauthor{\bsnm{Zhu}, \binits{N.}},
\bauthor{\bsnm{Xu}, \binits{K.}},
\bauthor{\bsnm{Dai}, \binits{Y.}}, \betal:
\batitle{Large-scale coherent {Ising} machine based on optoelectronic parametric oscillator}.
\bjtitle{Light: Science \& Applications}
\bvolume{11},
\bfpage{333}
(\byear{2022})
\end{barticle}
\endbibitem

%%% 21
\bibitem[\protect\citeauthoryear{B{\"o}hm et~al.}{2019}]{bohm2019poor}
\begin{barticle}
\bauthor{\bsnm{B{\"o}hm}, \binits{F.}},
\bauthor{\bsnm{Verschaffelt}, \binits{G.}},
\bauthor{\bsnm{Sande}, \binits{G.}}:
\batitle{A poor man’s coherent {Ising} machine based on opto-electronic feedback systems for solving optimization problems}.
\bjtitle{Nature Communications}
\bvolume{10},
\bfpage{3538}
(\byear{2019})
\end{barticle}
\endbibitem

%%% 22
\bibitem[\protect\citeauthoryear{Takata et~al.}{2016}]{takata201616}
\begin{barticle}
\bauthor{\bsnm{Takata}, \binits{K.}},
\bauthor{\bsnm{Marandi}, \binits{A.}},
\bauthor{\bsnm{Hamerly}, \binits{R.}},
\bauthor{\bsnm{Haribara}, \binits{Y.}},
\bauthor{\bsnm{Maruo}, \binits{D.}},
\bauthor{\bsnm{Tamate}, \binits{S.}},
\bauthor{\bsnm{Sakaguchi}, \binits{H.}},
\bauthor{\bsnm{Utsunomiya}, \binits{S.}},
\bauthor{\bsnm{Yamamoto}, \binits{Y.}}:
\batitle{A 16-bit coherent {Ising} machine for one-dimensional ring and cubic graph problems}.
\bjtitle{Scientific Reports}
\bvolume{6},
\bfpage{34089}
(\byear{2016})
\end{barticle}
\endbibitem

%%% 23
\bibitem[\protect\citeauthoryear{McMahon et~al.}{2016}]{mcmahon2016fully}
\begin{barticle}
\bauthor{\bsnm{McMahon}, \binits{P.L.}},
\bauthor{\bsnm{Marandi}, \binits{A.}},
\bauthor{\bsnm{Haribara}, \binits{Y.}},
\bauthor{\bsnm{Hamerly}, \binits{R.}},
\bauthor{\bsnm{Langrock}, \binits{C.}},
\bauthor{\bsnm{Tamate}, \binits{S.}},
\bauthor{\bsnm{Inagaki}, \binits{T.}},
\bauthor{\bsnm{Takesue}, \binits{H.}},
\bauthor{\bsnm{Utsunomiya}, \binits{S.}},
\bauthor{\bsnm{Aihara}, \binits{K.}}, \betal:
\batitle{A fully programmable 100-spin coherent {Ising} machine with all-to-all connections}.
\bjtitle{Science}
\bvolume{354},
\bfpage{614}--\blpage{617}
(\byear{2016})
\end{barticle}
\endbibitem

%%% 24
\bibitem[\protect\citeauthoryear{Inagaki et~al.}{2016}]{inagaki2016coherent}
\begin{barticle}
\bauthor{\bsnm{Inagaki}, \binits{T.}},
\bauthor{\bsnm{Haribara}, \binits{Y.}},
\bauthor{\bsnm{Igarashi}, \binits{K.}},
\bauthor{\bsnm{Sonobe}, \binits{T.}},
\bauthor{\bsnm{Tamate}, \binits{S.}},
\bauthor{\bsnm{Honjo}, \binits{T.}},
\bauthor{\bsnm{Marandi}, \binits{A.}},
\bauthor{\bsnm{McMahon}, \binits{P.L.}},
\bauthor{\bsnm{Umeki}, \binits{T.}},
\bauthor{\bsnm{Enbutsu}, \binits{K.}}, \betal:
\batitle{A coherent {Ising} machine for 2000-node optimization problems}.
\bjtitle{Science}
\bvolume{354},
\bfpage{603}--\blpage{606}
(\byear{2016})
\end{barticle}
\endbibitem

%%% 25
\bibitem[\protect\citeauthoryear{Babaeian et~al.}{2019}]{babaeian2019single}
\begin{barticle}
\bauthor{\bsnm{Babaeian}, \binits{M.}},
\bauthor{\bsnm{Nguyen}, \binits{D.T.}},
\bauthor{\bsnm{Demir}, \binits{V.}},
\bauthor{\bsnm{Akbulut}, \binits{M.}},
\bauthor{\bsnm{Blanche}, \binits{P.-A.}},
\bauthor{\bsnm{Kaneda}, \binits{Y.}},
\bauthor{\bsnm{Guha}, \binits{S.}},
\bauthor{\bsnm{Neifeld}, \binits{M.A.}},
\bauthor{\bsnm{Peyghambarian}, \binits{N.}}:
\batitle{A single shot coherent {Ising} machine based on a network of injection-locked multicore fiber lasers}.
\bjtitle{Nature Communications}
\bvolume{10},
\bfpage{3516}
(\byear{2019})
\end{barticle}
\endbibitem

%%% 26
\bibitem[\protect\citeauthoryear{Luo et~al.}{2023}]{WDM_Li2023}
\begin{barticle}
\bauthor{\bsnm{Luo}, \binits{L.}},
\bauthor{\bsnm{Mi}, \binits{Z.}},
\bauthor{\bsnm{Huang}, \binits{J.}},
\bauthor{\bsnm{Ruan}, \binits{Z.}}:
\batitle{Wavelength-division multiplexing optical {Ising} simulator enabling fully programmable spin couplings and external magnetic fields}.
\bjtitle{Science Advances}
\bvolume{9},
\bfpage{6238}
(\byear{2023})
\end{barticle}
\endbibitem

%%% 27
\bibitem[\protect\citeauthoryear{Ye et~al.}{2023}]{ye2023photonic}
\begin{barticle}
\bauthor{\bsnm{Ye}, \binits{X.}},
\bauthor{\bsnm{Zhang}, \binits{W.}},
\bauthor{\bsnm{Wang}, \binits{S.}},
\bauthor{\bsnm{Yang}, \binits{X.}},
\bauthor{\bsnm{He}, \binits{Z.}}:
\batitle{20736-node weighted max-cut problem solving by quadrature photonic spatial ising machine}.
\bjtitle{Science China Information Sciences}
\bvolume{66},
\bfpage{229301}
(\byear{2023})
\end{barticle}
\endbibitem

%%% 28
\bibitem[\protect\citeauthoryear{Ouyang et~al.}{2024}]{ouyang2024demand}
\begin{barticle}
\bauthor{\bsnm{Ouyang}, \binits{J.}},
\bauthor{\bsnm{Liao}, \binits{Y.}},
\bauthor{\bsnm{Ma}, \binits{Z.}},
\bauthor{\bsnm{Kong}, \binits{D.}},
\bauthor{\bsnm{Feng}, \binits{X.}},
\bauthor{\bsnm{Zhang}, \binits{X.}},
\bauthor{\bsnm{Dong}, \binits{X.}},
\bauthor{\bsnm{Cui}, \binits{K.}},
\bauthor{\bsnm{Liu}, \binits{F.}},
\bauthor{\bsnm{Zhang}, \binits{W.}}, \betal:
\batitle{On-demand photonic ising machine with simplified hamiltonian calculation by phase encoding and intensity detection}.
\bjtitle{Communications Physics}
\bvolume{7},
\bfpage{168}
(\byear{2024})
\end{barticle}
\endbibitem

%%% 29
\bibitem[\protect\citeauthoryear{Irb{\"a}ck et~al.}{2022}]{irback2022}
\begin{barticle}
\bauthor{\bsnm{Irb{\"a}ck}, \binits{A.}},
\bauthor{\bsnm{Knuthson}, \binits{L.}},
\bauthor{\bsnm{Mohanty}, \binits{S.}},
\bauthor{\bsnm{Peterson}, \binits{C.}}:
\batitle{Folding lattice proteins with quantum annealing}.
\bjtitle{Physical Review Research}
\bvolume{4},
\bfpage{043013}
(\byear{2022})
\end{barticle}
\endbibitem

%%% 30
\bibitem[\protect\citeauthoryear{Estar{\'a}n et~al.}{2019}]{estaran2019}
\begin{barticle}
\bauthor{\bsnm{Estar{\'a}n}, \binits{J.M.}},
\bauthor{\bsnm{Almonacil}, \binits{S.}},
\bauthor{\bsnm{Rios-M{\"u}ller}, \binits{R.}},
\bauthor{\bsnm{Mardoyan}, \binits{H.}},
\bauthor{\bsnm{Jenneve}, \binits{P.}},
\bauthor{\bsnm{Benyahya}, \binits{K.}},
\bauthor{\bsnm{Simonneau}, \binits{C.}},
\bauthor{\bsnm{Bigo}, \binits{S.}},
\bauthor{\bsnm{Renaudier}, \binits{J.}},
\bauthor{\bsnm{Charlet}, \binits{G.}}:
\batitle{{Sub-Baudrate} sampling at {DAC} and {ADC}: Toward 200{G} per lane {IM/DD} systems}.
\bjtitle{Journal of Lightwave Technology}
\bvolume{37},
\bfpage{1536}--\blpage{1542}
(\byear{2019})
\end{barticle}
\endbibitem

%%% 31
\bibitem[\protect\citeauthoryear{Wang et~al.}{2023}]{wang2023}
\begin{barticle}
\bauthor{\bsnm{Wang}, \binits{J.}},
\bauthor{\bsnm{Ebler}, \binits{D.}},
\bauthor{\bsnm{Wong}, \binits{K.M.}},
\bauthor{\bsnm{Hui}, \binits{D.S.W.}},
\bauthor{\bsnm{Sun}, \binits{J.}}:
\batitle{Bifurcation behaviors shape how continuous physical dynamics solves discrete {Ising} optimization}.
\bjtitle{Nature Communications}
\bvolume{14},
\bfpage{2510}
(\byear{2023})
\end{barticle}
\endbibitem

%%% 32
\bibitem[\protect\citeauthoryear{Zhang et~al.}{2022}]{zhang2022silicon}
\begin{barticle}
\bauthor{\bsnm{Zhang}, \binits{W.}},
\bauthor{\bsnm{Huang}, \binits{C.}},
\bauthor{\bsnm{Peng}, \binits{H.-T.}},
\bauthor{\bsnm{Bilodeau}, \binits{S.}},
\bauthor{\bsnm{Jha}, \binits{A.}},
\bauthor{\bsnm{Blow}, \binits{E.}},
\bauthor{\bsnm{De~Lima}, \binits{T.F.}},
\bauthor{\bsnm{Shastri}, \binits{B.J.}},
\bauthor{\bsnm{Prucnal}, \binits{P.}}:
\batitle{Silicon microring synapses enable photonic deep learning beyond 9-bit precision}.
\bjtitle{Optica}
\bvolume{9},
\bfpage{579}--\blpage{584}
(\byear{2022})
\end{barticle}
\endbibitem

%%% 33
\bibitem[\protect\citeauthoryear{B{\"o}hm et~al.}{2022}]{bohm2022noise}
\begin{barticle}
\bauthor{\bsnm{B{\"o}hm}, \binits{F.}},
\bauthor{\bsnm{Alonso-Urquijo}, \binits{D.}},
\bauthor{\bsnm{Verschaffelt}, \binits{G.}},
\bauthor{\bsnm{Sande}, \binits{G.}}:
\batitle{Noise-injected analog {Ising} machines enable ultrafast statistical sampling and machine learning}.
\bjtitle{Nature Communications}
\bvolume{13},
\bfpage{5847}
(\byear{2022})
\end{barticle}
\endbibitem

%%% 34
\bibitem[\protect\citeauthoryear{Benlic and Hao}{2013}]{benlic2013}
\begin{barticle}
\bauthor{\bsnm{Benlic}, \binits{U.}},
\bauthor{\bsnm{Hao}, \binits{J.K.}}:
\batitle{Breakout local search for the max-cut problem}.
\bjtitle{Engineering Applications of Artificial Intelligence}
\bvolume{26},
\bfpage{1162}--\blpage{1173}
(\byear{2013})
\end{barticle}
\endbibitem

%%% 35
\bibitem[\protect\citeauthoryear{Leleu et~al.}{2021}]{leleu2021}
\begin{barticle}
\bauthor{\bsnm{Leleu}, \binits{T.}},
\bauthor{\bsnm{Khoyratee}, \binits{F.}},
\bauthor{\bsnm{Levi}, \binits{T.}},
\bauthor{\bsnm{Hamerly}, \binits{R.}},
\bauthor{\bsnm{Kohno}, \binits{T.}},
\bauthor{\bsnm{Aihara}, \binits{K.}}:
\batitle{Scaling advantage of chaotic amplitude control for high-performance combinatorial optimization}.
\bjtitle{Communications Physics}
\bvolume{4},
\bfpage{266}
(\byear{2021})
\end{barticle}
\endbibitem

%%% 36
\bibitem[\protect\citeauthoryear{Kalinin et~al.}{2023}]{kalinin2023analog}
\begin{botherref}
\oauthor{\bsnm{Kalinin}, \binits{K.P.}},
\oauthor{\bsnm{Mourgias-Alexandris}, \binits{G.}},
\oauthor{\bsnm{Ballani}, \binits{H.}},
\oauthor{\bsnm{Berloff}, \binits{N.G.}},
\oauthor{\bsnm{Clegg}, \binits{J.H.}},
\oauthor{\bsnm{Cletheroe}, \binits{D.}},
\oauthor{\bsnm{Gkantsidis}, \binits{C.}},
\oauthor{\bsnm{Haller}, \binits{I.}},
\oauthor{\bsnm{Lyutsarev}, \binits{V.}},
\oauthor{\bsnm{Parmigiani}, \binits{F.}}, et al.:
Analog iterative machine {(AIM)}: using light to solve quadratic optimization problems with mixed variables.
arXiv preprint arXiv:2304.12594
(2023)
\end{botherref}
\endbibitem

%%% 37
\bibitem[\protect\citeauthoryear{Pramanik et~al.}{2024}]{pramanik2024}
\begin{bchapter}
\bauthor{\bsnm{Pramanik}, \binits{S.}},
\bauthor{\bsnm{Chatterjee}, \binits{S.}},
\bauthor{\bsnm{Oza}, \binits{H.}}:
\bctitle{Convergence analysis of opto-electronic oscillator based coherent Ising machines}.
In: \bbtitle{2024 16th International Conference on COMmunication Systems \& NETworkS (COMSNETS)},
pp. \bfpage{1076}--\blpage{1081}
(\byear{2024}).
%\bcomment{IEEE}
\end{bchapter}
\endbibitem

%%% 38
\bibitem[\protect\citeauthoryear{Abramson et~al.}{2024}]{abramson2024}
\begin{barticle}
\bauthor{\bsnm{Abramson}, \binits{J.}},
\bauthor{\bsnm{Adler}, \binits{J.}},
\bauthor{\bsnm{Dunger}, \binits{J.}},
\bauthor{\bsnm{Evans}, \binits{R.}},
\bauthor{\bsnm{Green}, \binits{T.}},
\bauthor{\bsnm{Pritzel}, \binits{A.}},
\bauthor{\bsnm{Ronneberger}, \binits{O.}},
\bauthor{\bsnm{Willmore}, \binits{L.}},
\bauthor{\bsnm{Ballard}, \binits{A.J.}},
\bauthor{\bsnm{Bambrick}, \binits{J.}}, \betal:
\batitle{Accurate structure prediction of biomolecular interactions with {AlphaFold} 3}.
\bjtitle{Nature}
\bvolume{630},
\bfpage{493}--\blpage{500}
(\byear{2024})
\end{barticle}
\endbibitem

%%% 39
\bibitem[\protect\citeauthoryear{Huang et~al.}{2021}]{huang2021}
\begin{barticle}
\bauthor{\bsnm{Huang}, \binits{J.}},
\bauthor{\bsnm{Fang}, \binits{Y.}},
\bauthor{\bsnm{Ruan}, \binits{Z.}}:
\batitle{Antiferromagnetic spatial photonic {Ising} machine through optoelectronic correlation computing}.
\bjtitle{Communications Physics}
\bvolume{4},
\bfpage{242}
(\byear{2021})
\end{barticle}
\endbibitem

%%% 40
\bibitem[\protect\citeauthoryear{Prabhakar et~al.}{2023}]{prabhakar2023}
\begin{barticle}
\bauthor{\bsnm{Prabhakar}, \binits{A.}},
\bauthor{\bsnm{Shah}, \binits{P.}},
\bauthor{\bsnm{Gautham}, \binits{U.}},
\bauthor{\bsnm{Natarajan}, \binits{V.}},
\bauthor{\bsnm{Ramesh}, \binits{V.}},
\bauthor{\bsnm{Chandrachoodan}, \binits{N.}},
\bauthor{\bsnm{Tayur}, \binits{S.}}:
\batitle{Optimization with photonic wave-based annealers}.
\bjtitle{Philosophical Transactions of the Royal Society A}
\bvolume{381},
\bfpage{20210409}
(\byear{2023})
\end{barticle}
\endbibitem

%%% 41
\bibitem[\protect\citeauthoryear{Asproni et~al.}{2020}]{asproni2020accuracy}
\begin{barticle}
\bauthor{\bsnm{Asproni}, \binits{L.}},
\bauthor{\bsnm{Caputo}, \binits{D.}},
\bauthor{\bsnm{Silva}, \binits{B.}},
\bauthor{\bsnm{Fazzi}, \binits{G.}},
\bauthor{\bsnm{Magagnini}, \binits{M.}}:
\batitle{Accuracy and minor embedding in subqubo decomposition with fully connected large problems: a case study about the number partitioning problem}.
\bjtitle{Quantum Machine Intelligence}
\bvolume{2},
\bfpage{4}
(\byear{2020})
\end{barticle}
\endbibitem

%%% 42
\bibitem[\protect\citeauthoryear{Xu et~al.}{2020}]{xu2020}
\begin{barticle}
\bauthor{\bsnm{Xu}, \binits{X.Y.}},
\bauthor{\bsnm{Huang}, \binits{X.L.}},
\bauthor{\bsnm{Li}, \binits{Z.M.}},
\bauthor{\bsnm{Gao}, \binits{J.}},
\bauthor{\bsnm{Jiao}, \binits{Z.Q.}},
\bauthor{\bsnm{Wang}, \binits{Y.}},
\bauthor{\bsnm{Ren}, \binits{R.J.}},
\bauthor{\bsnm{Zhang}, \binits{H.}},
\bauthor{\bsnm{Jin}, \binits{X.M.}}:
\batitle{A scalable photonic computer solving the subset sum problem}.
\bjtitle{Science Advances}
\bvolume{6},
\bfpage{5853}
(\byear{2020})
\end{barticle}
\endbibitem

%%% 43
\bibitem[\protect\citeauthoryear{Tariq et~al.}{2020}]{tariq2020computational}
\begin{barticle}
\bauthor{\bsnm{Tariq}, \binits{M.}},
\bauthor{\bsnm{Al-Dweik}, \binits{A.}},
\bauthor{\bsnm{Mohammad}, \binits{B.}},
\bauthor{\bsnm{Saleh}, \binits{H.}},
\bauthor{\bsnm{Stouraitis}, \binits{T.}}:
\batitle{Computational power evaluation for energy-constrained wireless communications systems}.
\bjtitle{IEEE Open Journal of the Communications Society}
\bvolume{1},
\bfpage{308}--\blpage{319}
(\byear{2020})
\end{barticle}
\endbibitem

%%% 44
\bibitem[\protect\citeauthoryear{Inagaki et~al.}{2016}]{inagaki2016}
\begin{barticle}
\bauthor{\bsnm{Inagaki}, \binits{T.}},
\bauthor{\bsnm{Haribara}, \binits{Y.}},
\bauthor{\bsnm{Igarashi}, \binits{K.}},
\bauthor{\bsnm{Sonobe}, \binits{T.}},
\bauthor{\bsnm{Tamate}, \binits{S.}},
\bauthor{\bsnm{Honjo}, \binits{T.}},
\bauthor{\bsnm{Marandi}, \binits{A.}},
\bauthor{\bsnm{McMahon}, \binits{P.L.}},
\bauthor{\bsnm{Umeki}, \binits{T.}},
\bauthor{\bsnm{Enbutsu}, \binits{K.}}, \betal:
\batitle{A coherent ising machine for 2000-node optimization problems}.
\bjtitle{Science}
\bvolume{354},
\bfpage{603}--\blpage{606}
(\byear{2016})
\end{barticle}
\endbibitem

%%% 45
\bibitem[\protect\citeauthoryear{Corporation}{2023}]{cienaWaveLogic}
\begin{botherref}
\oauthor{\bsnm{Corporation}, \binits{C.}}:
{I}ntroducing {W}ave{L}ogic 6: {A}nother industry first from {C}iena.
\url{https://www.ciena.com/insights/blog/2023/introducing-wavelogic-6?utm_source=blog&utm_medium=social}
(2023)
\end{botherref}
\endbibitem

%%% 46
\bibitem[\protect\citeauthoryear{Pappas et~al.}{2025}]{pappas2025}
\begin{botherref}
\oauthor{\bsnm{Pappas}, \binits{C.}},
\oauthor{\bsnm{Moschos}, \binits{T.}},
\oauthor{\bsnm{Prapas}, \binits{A.}},
\oauthor{\bsnm{Kirtas}, \binits{M.}},
\oauthor{\bsnm{Moralis-Pegios}, \binits{M.}},
\oauthor{\bsnm{Tsakyridis}, \binits{A.}},
\oauthor{\bsnm{Asimopoulos}, \binits{O.}},
\oauthor{\bsnm{Passalis}, \binits{N.}},
\oauthor{\bsnm{Tefas}, \binits{A.}},
\oauthor{\bsnm{Pleros}, \binits{N.}}:
Reaching the peta-computing: 163.8 {TOPS} through multidimensional {AWGR}-based accelerators.
Journal of Lightwave Technology
\bvolume{43},
\bfpage{1773}--\blpage{1785}
(2025)
\end{botherref}
\endbibitem

%%% 47
\bibitem[\protect\citeauthoryear{Bai et~al.}{2023}]{bai2023}
\begin{barticle}
\bauthor{\bsnm{Bai}, \binits{Y.}},
\bauthor{\bsnm{Xu}, \binits{X.}},
\bauthor{\bsnm{Tan}, \binits{M.}},
\bauthor{\bsnm{Sun}, \binits{Y.}},
\bauthor{\bsnm{Li}, \binits{Y.}},
\bauthor{\bsnm{Wu}, \binits{J.}},
\bauthor{\bsnm{Morandotti}, \binits{R.}},
\bauthor{\bsnm{Mitchell}, \binits{A.}},
\bauthor{\bsnm{Xu}, \binits{K.}},
\bauthor{\bsnm{Moss}, \binits{D.J.}}:
\batitle{Photonic multiplexing techniques for neuromorphic computing}.
\bjtitle{Nanophotonics}
\bvolume{12},
\bfpage{795}--\blpage{817}
(\byear{2023})
\end{barticle}
\endbibitem

%%% 48
\bibitem[\protect\citeauthoryear{NVIDIA}{2023}]{nvidiaNVIDIAH100Tensor}
\begin{botherref}
\oauthor{\bsnm{NVIDIA}}:
{{NVIDIA H100 Tensor Core GPU Architecture}}.
Whitepaper,
NVIDIA
(2023).
\url{https://resources.nvidia.com/en-us-tensor-core}
\end{botherref}
\endbibitem

%%% 49
\bibitem[\protect\citeauthoryear{Gill et~al.}{2024}]{gill2024quantum}
\begin{botherref}
\oauthor{\bsnm{Gill}, \binits{S.S.}},
\oauthor{\bsnm{Cetinkaya}, \binits{O.}},
\oauthor{\bsnm{Marrone}, \binits{S.}},
\oauthor{\bsnm{Claudino}, \binits{D.}},
\oauthor{\bsnm{Haunschild}, \binits{D.}},
\oauthor{\bsnm{Schlote}, \binits{L.}},
\oauthor{\bsnm{Wu}, \binits{H.}},
\oauthor{\bsnm{Ottaviani}, \binits{C.}},
\oauthor{\bsnm{Liu}, \binits{X.}},
\oauthor{\bsnm{Machupalli}, \binits{S.P.}}, et al.:
Quantum computing: Vision and challenges.
arXiv preprint arXiv:2403.02240
(2024)
\end{botherref}
\endbibitem

%%% 50
\bibitem[\protect\citeauthoryear{De~Leon et~al.}{2021}]{de2021materials}
\begin{barticle}
\bauthor{\bsnm{De~Leon}, \binits{N.P.}},
\bauthor{\bsnm{Itoh}, \binits{K.M.}},
\bauthor{\bsnm{Kim}, \binits{D.}},
\bauthor{\bsnm{Mehta}, \binits{K.K.}},
\bauthor{\bsnm{Northup}, \binits{T.E.}},
\bauthor{\bsnm{Paik}, \binits{H.}},
\bauthor{\bsnm{Palmer}, \binits{B.}},
\bauthor{\bsnm{Samarth}, \binits{N.}},
\bauthor{\bsnm{Sangtawesin}, \binits{S.}},
\bauthor{\bsnm{Steuerman}, \binits{D.W.}}:
\batitle{Materials challenges and opportunities for quantum computing hardware}.
\bjtitle{Science}
\bvolume{372},
\bfpage{2823}
(\byear{2021})
\end{barticle}
\endbibitem

\end{thebibliography}

\newpage

\backmatter

\section{Methods}
\label{Method}

\subsection{Mathematical framework}  \label{sec:methods_math_framework}
The iterative gradient descent on the Ising Hamiltonian
\begin{equation}
H = \sum_{i < j} J_{ij} \sigma_i \sigma_j + \sum_i h_i \sigma_i
\label{eq:ising_energy}
\end{equation}
is performed using continuous spins variables defined as $\sigma_i = \mathrm{sign}(x_i)$, where $x_i \in \mathbb{R}$. These are approximated using a saturating nonlinearity, such as a half-wave-bounded sinusoidal function. The state update at each iteration is governed by a moving average between the current state and the gradient of the energy function:
\begin{align}
\mathbf{x}^{(t+1)} &= \alpha \mathbf{x}^{(t)} - \beta \nabla_x H \nonumber \\
&\approx (\alpha \mathbf{I} - \beta \mathbf{J}) \cdot \sigma(\mathbf{x}^{(t)}) - \beta \mathbf{h}
\label{eq:state_update}
\end{align}
where the hyperparameters $\alpha$ and $\beta$ denote the feedback and coupling strength, respectively, and determine the convergence dynamics of the system. The state update equation can be expressed compactly as an MVM operation by concatenating the constant term $-\beta \mathbf{h}$ onto the matrix $\mathbf{W}=\alpha \mathbf{I} - \beta \mathbf{J}$ and appending a value of 1 onto $\mathbf{x}$:
\begin{align}
\mathbf{W}&=[\alpha \mathbf{I} - \beta \mathbf{J} | -\beta\mathbf{h}] \quad\text{and}\quad \mathbf{x'}^T=[\mathbf{x}^T | 1],\
\mathbf{x}^{(t+1)} = \mathbf{W} \cdot \sigma(\mathbf{x'}^{(t)}) + \eta(\zeta)
\label{eqn:augment_W_X}
\end{align}
where $\eta(\zeta) \sim \mathcal{N}(0, \zeta^2)$ is a normally distributed noise term introducing stochasticity, parameterized by variance $\zeta^2$. Following this framework, any computational task that can be formulated as energy minimization may be heuristically solved via iterative MVM with nonlinear feedback. The task is reduced to determining appropriate values for $\mathbf{J}$, $\mathbf{h}$, optimizing the hyperparameters $\alpha$ and $\beta$, and the annealing schedule for $\zeta^2$.

\subsubsection*{Time-multiplexed matrix-vector multiplication}
For dense MVM, time-multiplexed multiplication involves flattening the matrix $\mathbf{W}$ into a vector $\Tilde{\mathbf{w}}$, and repeating the spin vector $\mathbf{x'}$, $N$ times to form a corresponding vector $\Tilde{\mathbf{x}}$ of equal length as $\Tilde{\mathbf{w}}$. Element-wise multiplication is then performed between $\Tilde{\mathbf{w}}$ and $\Tilde{\mathbf{x}}$, followed by summation. These vectors can be converted into time-domain signals \( \tilde{x}(t) \) and \( \tilde{w}(t) \) via 

\begin{equation}
\tilde{x}(t) = \sum_{j=1}^{N(N+1)} \int_{0}^{\infty} \tilde{x}_j \delta(t - \frac{j}{f_s})dt
\end{equation}

\begin{equation}
\tilde{w}(t) = \sum_{j=1}^{N(N+1)} \int_{0}^{\infty} \tilde{w}_j \delta(t - \frac{j}{f_s}+\Delta T) dt
\end{equation} 
where $\delta$ is the Dirac delta function, $f_s$ is the baud rate, and $\Delta T$ is the travel time of the light between the two MZMs. This timing offset ensures correct element-wise time-multiplexed multiplication of $\Tilde{\mathbf{x}}$ and $\Tilde{\mathbf{w}}$. The summation is performed up to $j=N(N+1)$ because of the change in shape described in equation~(\ref{eqn:augment_W_X}). For sparse matrices, the same procedure applies, except a preprocessing step determines the positions of the zero elements of $\mathbf{W}$, and removes corresponding matrix and vector elements during the flattening. These signals $\Tilde{x}(t)$ and $\Tilde{w}(t)$ are applied as drive voltages from the DACs to two MZMs. The transfer functions of the MZMs then perform analog multiplication in the optical domain. In our implementation, two cascaded MZMs are followed by a PD. The resulting output voltage is:

\begin{equation}
V_{\text{out}}(t) =  I_0 \mathcal{R} \cos^2 \left(\frac{\pi}{4} \left( \frac{V_{1}}{V_\pi} \Tilde{x}(t) - 1 \right)\right) \cos^2 \left(\frac{\pi}{4} \left( \frac{V_{2}}{V_\pi} \Tilde{w}(t) - 1 \right)\right)
\label{eqn:cascaded_mzms}
\end{equation}
where $\mathcal{R}$ is the PD responsivity, V$_1$ and V$_2$ are the drive voltage amplitudes, and $V_\pi$ is the MZM's half-wave voltage. We can write the driving signals as $w(t) \approx \dfrac{\pi}{2} \dfrac{V_{2}}{V_\pi} \Tilde{w}(t)  $ and  $x(t) \approx \dfrac{\pi}{2} \dfrac{V_{1}}{V_\pi} \Tilde{x}(t)$, and noting that the first MZM operates in the nonlinear region while the second MZM in the linear regime, Equation~(\ref{eqn:cascaded_mzms}) simplifies to:

\begin{align}
V_{\text{out}}(t) &= \frac{I_0 \mathcal{R}}{4} \left( 1  +  \sin( x(t)) +  w(t) +  w(t) \sin (x(t)) \right) 
\end{align}
To extract the desired element-wise time multiplexed output $\mathbf{y}= \tilde{\mathbf{w}} \odot \sin \tilde{\mathbf{x}}$, one  can use a balanced photodetection or a combination of DC filtering and a time-interleaving encoding scheme~\cite{lin2024}, yielding:

\begin{align}
y(t) &=  \frac{I_0\mathcal{R}}{4} w(t) \sin (x(t))
\end{align}
The final MVM result is obtained from the resultant signal $y(t)$ by summing over the range that covers a given index. The summation can be performed digitally after ADC sampling or in the analog domain via an integrator before sampling, resulting in:

\begin{equation}
y_i = \sum_{j=1}^{N+1} W_{ij} \sigma(x_j')\text{.}
\end{equation}

\subsubsection*{Max-Cut}
In computer science, the Max-Cut problem involves partitioning the vertices \(V\) of a weighted graph \( G = (V, E) \) into two subsets \(S_+\) and \(S_-\), such that the sum of the weights of the edges \(E\) connecting the two subsets is maximized. For a given graph with edge weights \( k_{ij} \) between vertices \(i\) and \(j\), the cut value \( C \) can be defined as:

\begin{equation}
    C = \sum_{(i,j) \in E} k_{ij} \frac{1 - \sigma_i \sigma_j}{2}
    \label{eqn:cut_value}
\end{equation}
where \( \sigma_i \) is the spin variable associated with vertex \(i\), which takes values \( \pm 1 \) indicating the subset \(S_+\) or \(S_-\) to which it belongs to. The term \( (1 - \sigma_i \sigma_j)/2 \) equals \( 1 \) if vertices \( i \) and \( j \) belong to different subsets (i.e., the edge is cut), and \( 0 \) if they belong to the same subset (i.e., the non-cut edge). We define the corresponding Max-Cut Hamiltonian as:

\begin{align}
    H&=W-C \nonumber\\
    &=\sum_{(i,j) \in E} \frac{k_{ij}\sigma_i \sigma_j}{2} 
    \label{eqn:max_cut_hamiltonian}
\end{align}
where the sum of all the edge weights \( W = \sum_{(i,j) \in E} k_{ij}/2 \) offsets the constant term of $C$. Since \(W\) is fixed for a given graph, minimizing the Hamiltonian \(H\) is equivalent to maximizing the cut value $C$. Comparing Equation~\ref {eq:ising_energy}) with the general Ising energy function (Equation~\ref{eqn:max_cut_hamiltonian}) shows that the Max-Cut problem can be mapped to an Ising model by setting the spin coupling values as:
\begin{equation}
    J_{ij}=\frac{k_{ij}}{2}
\end{equation}

\subsubsection*{Folding lattice proteins}
The HP lattice protein folding problem is a simplified model of the physical process of protein folding that uses only two types of amino acids—hydrophobic (H) and polar (P)—placed on a two-dimensional grid. While multiple formulations of this problem exist, we follow the quadratic unconstrained binary optimization (QUBO) approach introduced in~\cite{irback2022}. The energy function in this formulation is given by:

\begin{equation}
    H = \mathbf{b}^T \mathbf{Q} \mathbf{b}
    \label{eqn:qubo_hamiltonian}
\end{equation}
where $b_i\in\{0,1\}$ are binary variables, and the index $i$ spans the product space of all combinations of amino acid sequence index and lattice site indices. Each variable $b_i$ indicates whether a particular amino acid is present at a given lattice site. The interaction values $Q_{ij}$ encode: (1) the energy reduction due to adjacent H--H interactions, and (2) the constraints that enforce valid chain configurations. For further details on the construction of the matrix $\mathbf{Q}$ for a given amino acid sequence, refer to~\cite{irback2022}. QUBO problems can be mapped to Ising problems via a one-to-one transformation between the binary variables $b_i$ and Ising spins  $\sigma_i$, using the mapping $\sigma_i=2b_i-1$. Substituting this relationship into Equation~(\ref{eqn:qubo_hamiltonian}), and comparing it with the Ising Hamiltonian in Equation~(\ref{eq:ising_energy}) yields the following expressions for the Ising coupling matrix $\mathbf{J}$ and local fields $\mathbf{h}$ in terms of $\mathbf{Q}$:

\begin{align}
    J_{ij} &=\frac{Q_{ij}}{4} \\
    h_i &= \frac{Q_{ii}}{2} + \sum_j \frac{Q_{ij}}{4}
\end{align}

Note that this transformation introduces a constant energy offset between the QUBO and Ising formulations.  However, since our solver is based on gradient descent, only the relative energies between configurations (i.e., the gradient) are relevant, and this constant offset has no effect on the final solution.

\subsubsection*{Number partitioning} \label{sec:methods_number_partitioning}
The objective of the number partitioning problem is to partition a set \( S=\{s_1, s_2, \dots, s_N\} \) into two subsets \( S_+, S_- \subseteq S \), such that the sums of the elements in each subset are equal~\cite{lucas2014ising}. In our experiment, we randomly select a set of $N$ positive integers from the range $\{x \mid x \in \mathbb{Z}, 0 \leq x \leq 16\}$. By assigning a spin value $\sigma_i \in \{\pm1\}$ to each number $s_i \in S$, the difference between the two subset sums can be written as:
\begin{equation}
    \Delta =  \sum_{s_i \in S_+}s_i -  \sum_{s_i \in S_-}s_i = \sum_{i=1}^N s_i \sigma_i
\end{equation}

Partitioning $S$ thus reduces to minimizing the square of this difference, yielding the number partitioning Hamiltonian:
\begin{align}
    H &= \Delta^2 - \sum_{s_i \in S} s_i^2 \nonumber \\
    &= \left(\sum_{i=1}^N s_i \sigma_i \right)^2 - \sum_{i=1}^N s_i^2 \nonumber \\
    &= \sum_{i,j | i \neq j}^N s_i s_j \sigma_i \sigma_j
    \label{eqn:partitioning_hamiltonian}
\end{align}

As with the Max-Cut problem, the constant term in the Hamiltonian is omitted, so the result depends only on the interaction term. Minimizing this Hamiltonian corresponds to minimizing the difference between the subset sums. From Equation~(\ref{eqn:partitioning_hamiltonian}), the Ising coupling coefficients are:
\begin{equation}
    J_{ij} = s_i s_j
\end{equation}

\subsection{Experimental testbed} \label{expSetupSupp}
The experimental setup of the cascaded modulator Ising machine (CMIM) used in our demonstrations, along with the digital signal processing (DSP) algorithms employed in the digital feedback path, is shown in Extended Data Fig.~\ref{fig:expSetup}. Analog electrical signals were generated using a two-channel Keysight M8199B arbitrary waveform generator (AWG) with a sampling rate of 256~GSa/s. Channel~1 encoded the $N$ Ising spins (repeated $N$ times), while Channel~2 encoded the fixed weight matrix (flattened and serialized). Both signals were streamed sequentially in the time domain to drive the thin-film lithium niobate (TFLN) modulators and perform element-wise multiplications as part of the matrix-vector multiplication (MVM) for the Ising problem under consideration. The first TFLN Mach–Zehnder modulator (MZM) converted the electrical signal carrying the $N$ spin amplitudes ($\mathbf{x}$) into an optical signal. The second TFLN MZM applied the weight matrix coefficients ($\mathbf{W}$). The relative delay between the two AWG channels was finely adjusted in software to ensure correct temporal alignment of $\mathbf{X}$ and $\mathbf{W}$. Channel~2 ($\mathbf{W}$) was driven with an RF voltage that ensured linear operation of the MZM, while Channel~1 ($\mathbf{X}$) used a higher RF amplitude to span the full nonlinear half-wave of the MZM’s electro-optic response. The TFLN modulators had a measured 3~dB bandwidth of 110~GHz and DC half-wave voltage $V_\pi=1.5$~V. On the transmission side, the DSP pipeline prepared the spins of the current iteration and corresponding weights for modulation. Pilot symbols (8,192 alternating ones and zeros in all experiments) were prepended to the spin vector up to the maximum number of symbols. The combined pilot and spin data sequence was then upsampled to 4 samples per symbol, and shaped using a root-raised cosine (RRC) filter followed by a pre-emphasis filter to compensate for known transmission-path losses. Finally, the signal was upsampled to generate the AWG modulation waveform. The Ising algorithm hyperparameters ($\alpha$ and $\beta$) and the RRC roll-off factors used for each experiment are listed in Extended Data Table~\ref{tab:hyperparameters}.

After travelling through the cascaded modulators and being detected by a photodetector (PD) with 100~GHz 3~dB bandwidth, the resulting signal was sampled with a 256~GSa/s Keysight UXR real-time oscilloscope (RTO). After sampling the waveform, the receiver-side DSP pipeline was applied: the mean was removed, the signal was resampled to 2 samples per symbol, and synchronization was performed using the known transmitted pilot sequence to correctly align spin indices. A 51-tap feed-forward equalizer was trained using only the pilot symbols, and the optimized filter was then convoluted with the received sequence. Finally, the signal was resampled to 1 sample per symbol, and the pilot symbols were discarded. The samples represented the element-wise multiplication results, and a summation operation was performed across the samples to produce a single value for each spin to be used in the following iteration. System performance was influenced by multiple noise sources, including laser noise, SOA noise, thermal noise, distortion from the AWG and PD, quantization noise, and DSP-induced inter-symbol interference. To emulate annealing and help the system escape local minima, controlled optical noise was injected via a bulk semiconductor optical amplifier (SOA), further described in Section~\ref{sec:optical_noise}. The optical signal and injected noise were combined using a 50:50 fiber coupler and amplified by a quantum dot semiconductor optical amplifier (QD SOA).

\subsection{Controllable optical noise source for annealing}
\label{sec:optical_noise}
To overcome local minima and improve convergence, the Ising machine algorithm requires controlled noise injection---either through constant noise at variable levels or a noise schedule that decreases over iterations, depending on the task. We investigated using a bulk SOA to inject controllable optical noise directly into the system. A bulk SOA generates amplified spontaneous emission (ASE) noise, which introduces uncertainty into the received optical signal. As a Poisson process with a high number of events, the distribution of the noise is roughly Gaussian. Extended Data Fig.~\ref{fig:noiseExperiment} summarizes our experimental findings. We first characterized how the bias current of the SOA affects the injected noise. The DFB laser and AWG modulation were disabled, and the received waveform was sampled by the RTO across a range of bias currents from 0 to 110~mA. Extended Data Fig.~\ref{fig:noiseExperiment}(b) shows the distribution of the received waveform for each current, which is approximately Gaussian, with a variance increasing with current. Extended Data Fig.~\ref{fig:noiseExperiment}(a) plots the corresponding noise variance and total measured noise power as functions of the bias current. Next, we examined how a modulated waveform is degraded by optical noise. For each SOA bias current, a signal representative of a typical task was modulated, and the optical spectrum of the output was measured using an optical spectrum analyzer. Extended Data Fig.~\ref{fig:noiseExperiment}(c) shows the optical signal-to-noise ratio (OSNR) computed as the ratio of the signal power to the noise power. We then tested the performance impact of this noise source in a 400-node (20$\times$20) 2D square lattice task, operating at 64~GBaud. The experiment was repeated at various SOA bias currents, each corresponding to a different constant noise level. Extended Data Fig.~\ref{fig:noiseExperiment}(d) shows the number of iterations to reach the ground state for each current. Each point reflects the mean of 10 trials. A clear minimum in iteration count is observed at 60~mA, indicating an optimal noise level for convergence. Finally, we implemented an exponential noise schedule, comparing it to constant-noise results. Using the noise characterization data, we selected bias currents such that the noise variance at iteration $t$, denoted $\eta_t^2$ follows the exponential annealing schedule $\eta_t^2=\eta_0^2\exp(-\gamma t)$. We experimentally determined optimal parameters: $\eta_0^2=18.2~\text{mV}^2$ and $\gamma=0.0235~\text{iter}^{-1}$. Extended Data Fig.~\ref{fig:noiseExperiment}(e) shows the number of iterations to reach the ground state comparing three cases: no optical noise (145$\pm$15 iterations); optimal constant noise at 60~mA (120$\pm$30 iterations); and exponential schedule (100$\pm$25 iterations). These results demonstrate that optical noise injection, particularly with a well-tuned annealing schedule, can significantly accelerate convergence to the ground state. 

\subsection{D-Wave quantum annealer benchmarking}

The latest D-Wave QPU processor, \textit{Advantage 4.1}, features 5,000 qubits and over 15,000 couplers. It operates at cryogenic temperatures of approximately 15 millikelvin (mK) to enable quantum annealing, with annealing times ranging from 0.5 $\mu$s to 2000 $\mu$s. The system employs a Pegasus topology, which provides a maximum qubit connectivity of 15 and allows problems to be embedded onto the hardware by mapping them onto this fixed graph structure. Since the number partitioning problem requires all-to-all connectivity, embedding onto D-Wave's architecture necessitates additional preprocessing via native clique embedding. To address this, we employed the \texttt{DWaveClique} solver, which is optimized for dense binary quadratic models with all-to-all connectivity. Using this approach, \textit{Advantage 4.1} can support up to 177 fully connected spins. For our experiments, the number partitioning problem was formulated as an Ising model, following the mapping described in 
Section~\ref{sec:methods_number_partitioning}. We then used the native clique embedding to map the resulting Ising model onto the Pegasus hardware graph. To estimate the success probability of finding the ground state, defined as
\begin{equation}
P_{\text{gs}} = \frac{\text{Number of successful runs}}{\text{Total number of runs}},
\end{equation}
we conducted experiments across problem sizes ranging from $N = 4$ to $N = 32$ in increments of 4. To ensure a fair scaling analysis, we specifically selected problem instances that had only one unique solution. Each problem size was executed $10^4$ times to compute average success rates. All experiments were conducted using D-Wave's Leap quantum cloud service, which provides real-time remote access to the QPU.

\bigskip

\bmhead{Data availability} The data presented in this work is available at: \href{https://github.com/Shastri-Lab/tfln-ising-nature-paper-2025}{https://github.com/tfln-ising-paper-2025}
\bmhead{Code availability} The experimental code  supporting the findings of this work is  available at: \href{https://github.com/Shastri-Lab/tfln-ising-nature-paper-2025}{https://github.com/tfln-ising-paper-2025}

\bmhead{Acknowledgements}
This work was supported in part by the Natural Sciences and Engineering Council of Canada (NSERC), Canada  Foundation for Innovation (CFI), and Ontario Research Funding. BJS and DVP are each supported by the Canada Research Chairs Program. HyperLight Corporation provided the thin-film lithium niobate modulators. The quantum-dot semiconductor optical amplifier was provided by Innolume GmbH. Access to the Advantage 4.1 quantum processing unit (QPU) was provided by D-Wave Quantum Inc. through their Leap real-time quantum cloud service.

\bmhead {Author contribution} 
BJS, NAK, HM and AA conceived the idea of the Ising architecture, developed the algorithms and theoretical framework. CSA and DVP conceived the idea of the DSP engine, developed the DSP algorithms, and built the experimental testbed. NAK, AA, CSA and HM designed the experiments, with input from BJS and DVP. CSA and NAK performed the experiments. HM and NAK mapped the Ising model algorithms and carried out numerical simulations. All authors discussed and analyzed the data. AA led the writing, with contributions from NAK, HM, CSA, DVP, and BJS. DVP and BJS supervised the study.

\bmhead{Conflict of interest/Competing interests} 
NAK, CSA, HM, AA, DVP, and BJS are involved in developing technologies at Milkshake Technology Inc.

\bmhead{Additional information}
\bmhead{Supplementary information} 
The online version contains supplementary material available at https://doi.org/xxx/xxx. \bmhead{Correspondence and requests for materials} Should be addressed to Bhavin J. Shastri.

\bigskip

\newpage
\setcounter{figure}{0}
\renewcommand{\figurename}{Extended Data Fig.}
\setcounter{table}{0}
\renewcommand{\tablename}{Extended Data Table}

\begin{figure}[htbp]
\centering
\includegraphics[width =5 in]{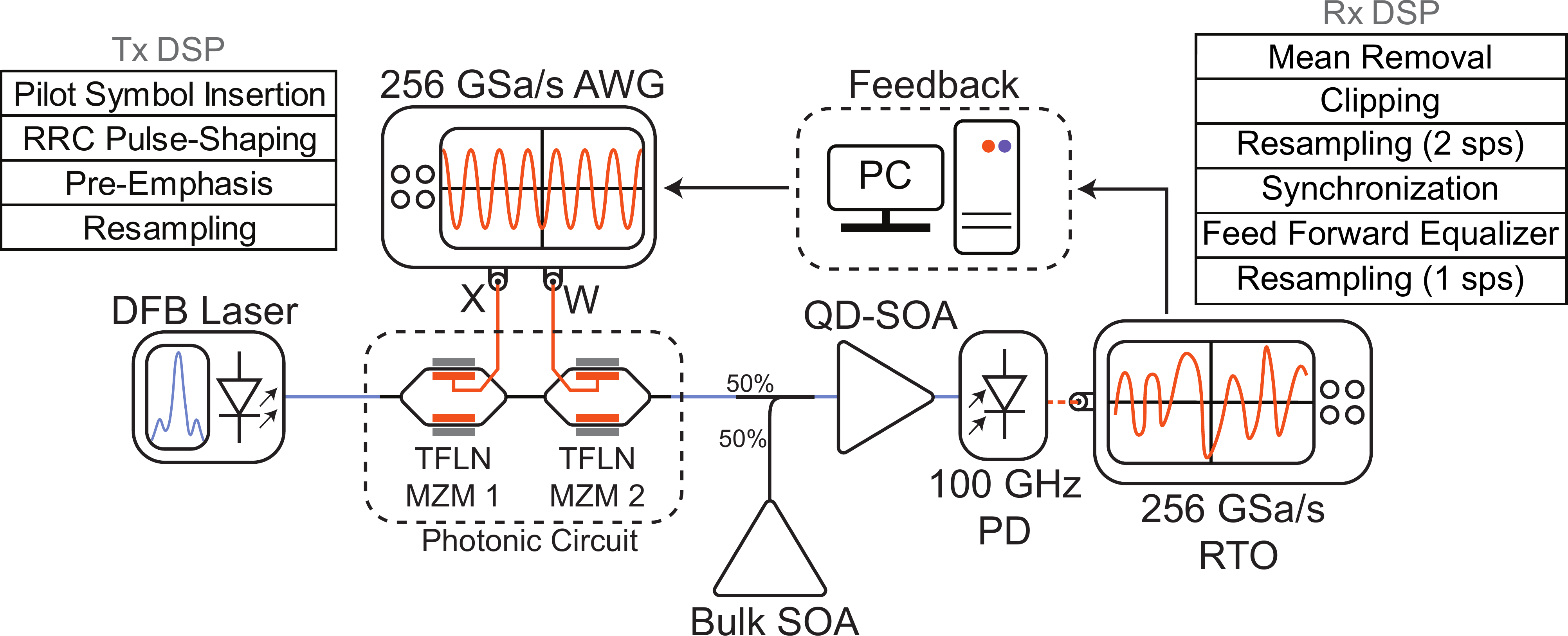}

 \caption{\textbf{Detailed experimental schematic of the CMIM.}. The electrical signals were generated using a two-channel Keysight M8199B arbitrary waveform generator (AWG) as the DAC with a sampling rate of 256 GSa/s. Channel~1 encoded the Ising spins, while Channel~2 encoded the fixed weight matrix. The TFLN modulators have a measured 3-dB bandwidth of 110 GHz and a DC half-wave voltage of $V_\pi=1.5$ V. The optimized transmitter (Tx) and receiver (Rx) digital signal processing (DSP) stack have been employed during high-speed operation. The modulated optical
signal was detected using a PD with a 3dB bandwidth of 100 GHz, and sampled by a 256 GSa/s Keysight UXR real-time oscilloscope (RTO) as the ADC. }

\label{fig:expSetup}
\end{figure}

\newpage
\begin{figure}[ht]
\centering
\includegraphics[width = 6 in]{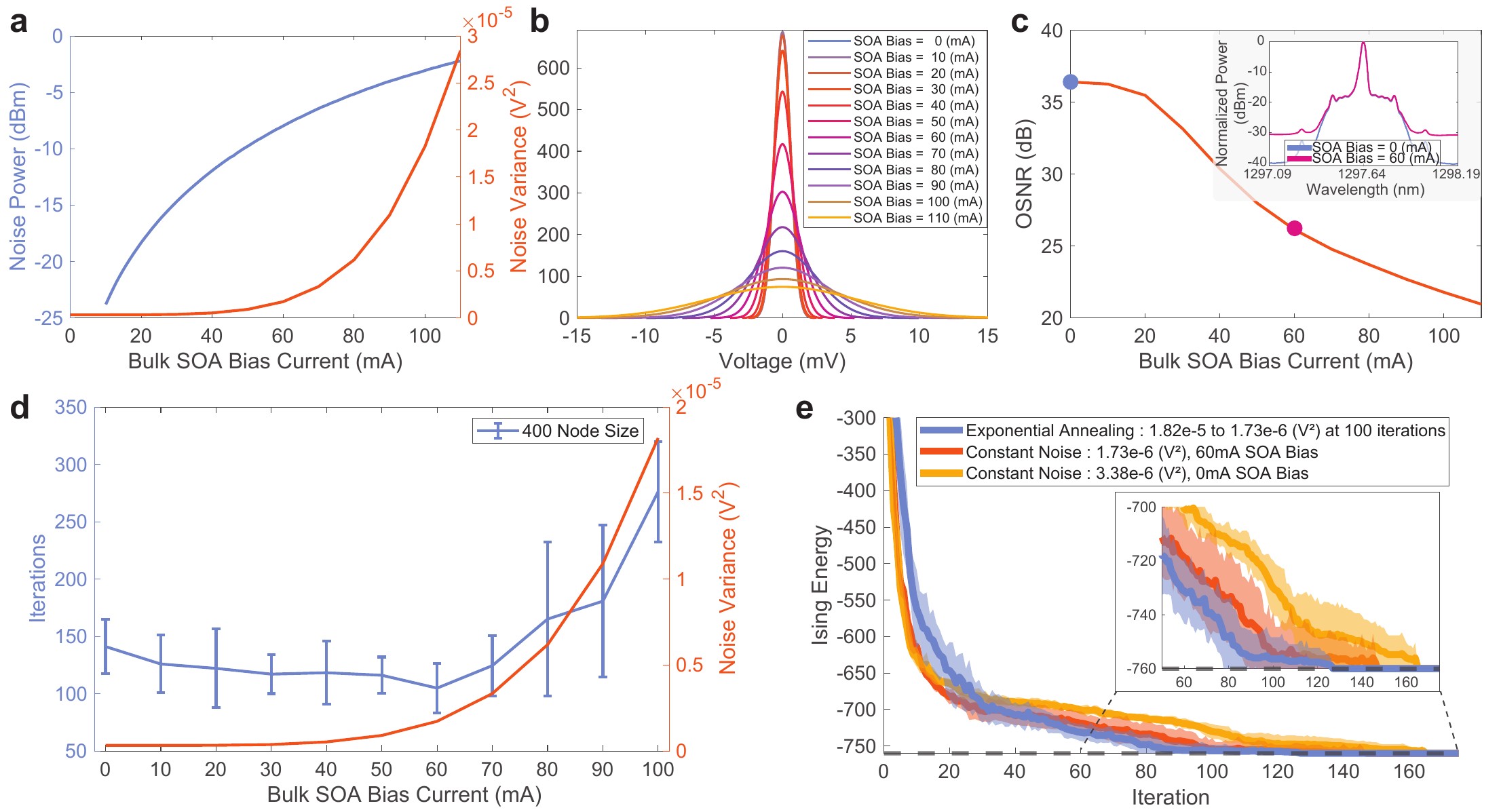}
 \caption{\textbf{Optical noise injection with a bulk SOA.} (a) ASE output power directly out of the bulk SOA and noise variance after measuring with noise with an RTO according to the schematic shown in Fig. \ref{fig:expSetup}. (b) Fit Gaussian probability density functions of the measured noise at the RTO for different SOA bias currents. (c) Measured OSNR at varying bulk SOA bias currents; waveform is data modulated for a 400-node lattice problem. The inset shows the wavelength spectrum for two different SOA biases. (d) Iterations required to reach ground state for a 400-node 2D lattice problem, depending on the bulk SOA bias current running at 64~Gbaud. (e) Evolution of the energy of a 400-node 2D lattice problem with constant noise (0~mA and 60~mA bulk SOA bias) and exponential annealing (initial SOA bias is 100 mA).}
\label{fig:noiseExperiment}
\end{figure}

\newpage
\begin{table}[h!]
\centering
\caption{Optimized algorithm hyperparameters for each demonstrated experiment operating at 106~GBaud. Sparse connections are the number of nonzero elements in the coupling weight matrix $\mathbf{W}$. $\alpha$ and $\beta$ are the hyperparameters used in the Ising algorithm. Roll-off is the beta factor of the RRC pulse-shaping.}
\label{tab:hyperparameters}
\renewcommand{\arraystretch}{1.5}
\begin{tabular}{ccccccccc}
    \hline
      \textbf{Tasks}  & \textbf{Spins (N)} & \textbf{Sparse connections} & & \textbf{$\alpha$} & \textbf{$\beta$}  & \textbf{Roll-Off} \\
    \hline 
    Square Lattice (101$\times$101) & 10,201 &  50,601 &  & 0.86 & 1   & 0.2  \\
    Square Lattice (203$\times$203) & 41,209 & 205,233  & & 0.86 & 1  & 0.2 \\
    Max-Cut (G22 Graph) & 2,000 &  41,980 & & 0.7 & 0.78 & 0.4  \\
    Max-Cut (G81 Graph) & 20,000 & 100,000 & & 0.7 & 0.78   & 0.4  \\
    Number Partitioning & 256 & 65,536 & & 1  & 0.3  & 0.2 \\
    Protein Folding (S30) & 630 & 31,920 & & 1  & 0.04  & 0.2  \\
    \hline 
\end{tabular}

\end{table}

\end{document}